\documentclass[11pt,a4paper]{article}%
\usepackage{amsmath,amssymb}%
\usepackage{geometry}%
\usepackage{graphicx}%
\usepackage{indentfirst} 
\usepackage[usenames]{color}
\begin{document}

\title{Dynamical properties in uniform and periodic growth modes
of ascorbic acid crystal domain from thin solution film}


\author{Yoshihiro Yamazaki, Mitsunobu Kikuchi, Akihiko Toda$^{1}$,
Jun-ichi Wakita$^{2}$, Mitsugu Matsushita$^{2}$\\
\ \\
Department of Physics, Waseda University, Ohkubo, Shinjuku-ku, Tokyo 169-8555\\
$^{1}$ Graduate School of Integrated Arts and Sciences, Hiroshima University,\\
Kagamiyama, Higashi-Hiroshima 739-8521\\
$^{2}$ Department of Physics, Chuo University,
Kasuga, Bunkyo-ku, Tokyo 112-8551
}

\maketitle

\begin{abstract}%
There exists the threshold-sensitive dynamical transition 
between the uniform and the periodic growth modes 
in the domain growth of ascorbic acid crystals 
from its aqueous supersaturated solution film.
The crystal growth induces the solution flow.
Humidity controls the fluidity of the solution.
The solution flow varies the film thickness.
The threshold exists in the thickness of the solution film, 
and the crystal growth almost stops
if the thickness becomes lower than the threshold.


\end{abstract}%

\\

\baselineskip 28pt

\section{Introduction}
\label{sec:intro}


Periodic growth from a point source causes concentric ring patterns,
although their growth mechanisms are essentially different.
%
They can be observed in a wide variety of systems such as
Belousov-Zhabotinsky reaction \cite{Epstein96, Epstein98}, 
bacterial colony growth \cite{Itoh99, Wakita01}, 
Liesegang precipitation \cite{Henisch05, Terada48},
and spherulite growth from a polymeric solution \cite{Wang07, Wang08} 
or from a solution
of an organic compound\cite{Gunn09, Shtukenberg11, Shtukenberg12} 
including ascorbic acid as one example.

\begin{figure}[!p]
\begin{center}
\includegraphics*[width=8cm]{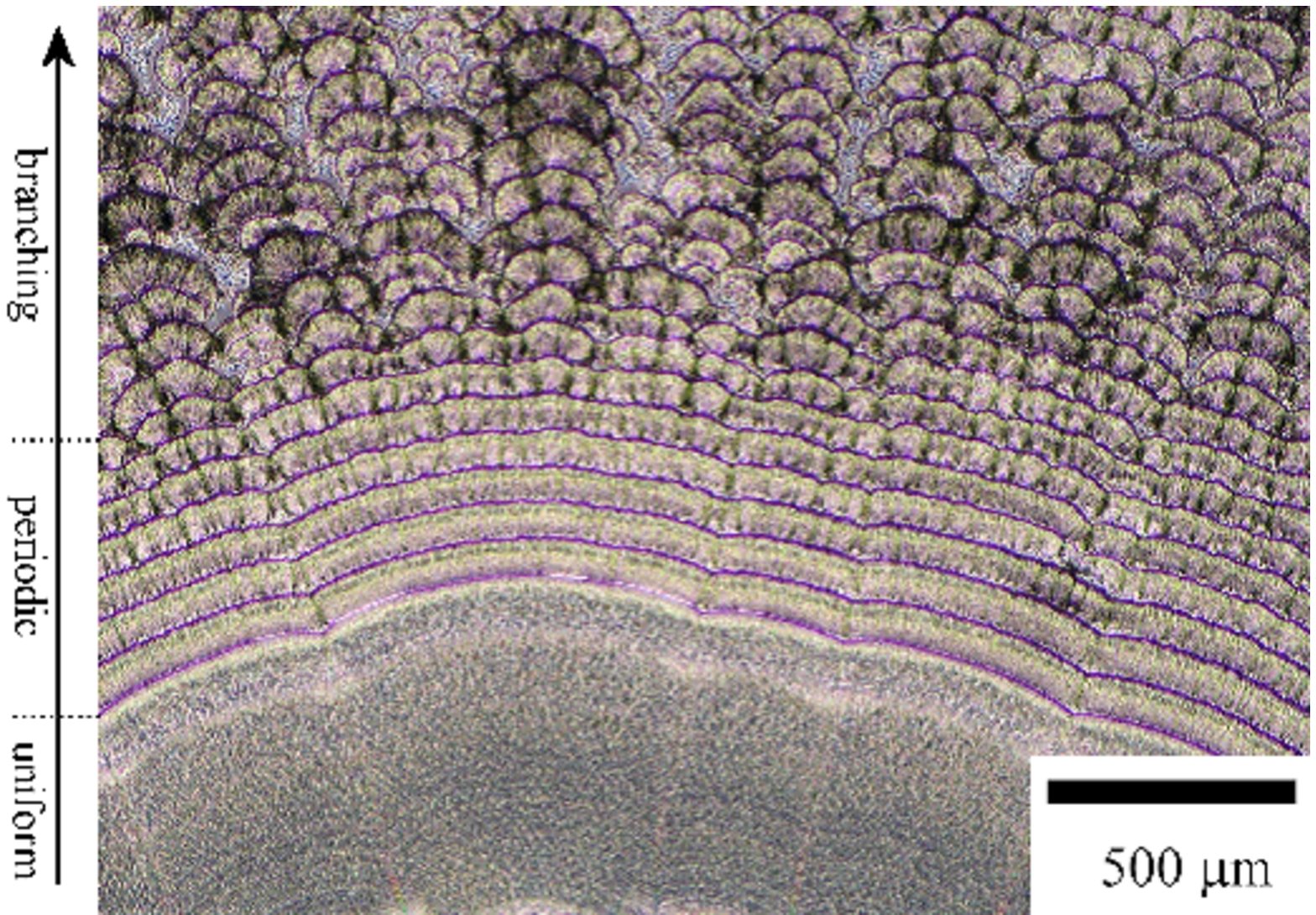}
\caption{Variation of domain patterns formed by change 
of the growth modes in gradual increase of humidity 
($\rho =$ 0.5 mg/cm$^{2}$, $T =$ 28 $^{\circ}$C).
The black arrow shows the growth direction.}
\label{fig:grad}
\end{center}
\end{figure}

\begin{figure}[!p]
\begin{center}
\includegraphics*[width=8cm]{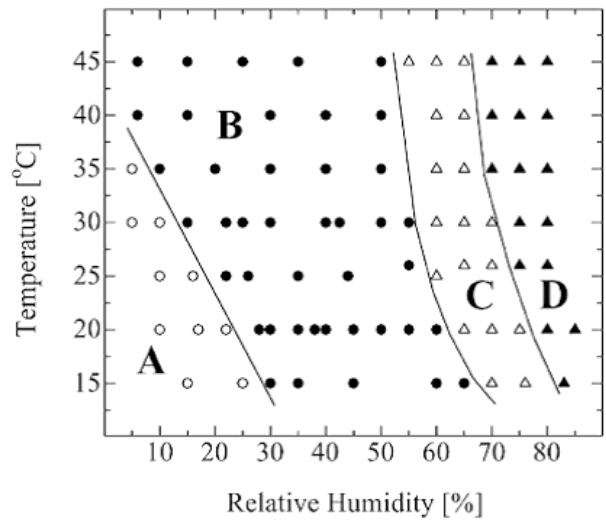}
\caption{Morphology diagram reprinted from Ref.\cite{Yamazaki09}
($\rho \approx$ 1.0 mg/cm$^{2}$) 
A: coexistence, B: uniform, C: periodic, and D: branching.}
\label{fig:diagram}
\end{center}
\end{figure}

%
For concentric ring pattern formation
in crystal growth of ascorbic acid from its thin solution film,
there have been several experimental studies
(from methanol solution\cite{Iwamoto84, Fukunaga96, Uesaka02, Uesaka03, Ito03}
and from aqueous solution\cite{Paranjpe02, Yamazaki09}).
%
The findings in these experimental studies
are summarized as follows.
%
The solution becomes a supersaturated state by solvent evaporation
before starting crystal growth.
%
Fluidity of the supersaturated solution 
is quite low especially in a low humidity condition,
so that the solution is expressed
as ``glassy''\cite{Gunn09}, ``amorphous''\cite{Uesaka02, Uesaka03}, 
and ``soft-solid''\cite{Paranjpe02}.
%
After that, ascorbic acid crystals begin to grow with nucleation.
%
A domain occupied with the ascorbic acid crystals spreads in (2+1) dimensions
because of growth from the thin solution film.
%
Namely, the domains can be considered as a quasi two dimensional pattern
with height.
%
The interesting point is that growth mode of the crystal domain
changes depending on environmental humidity.
%
With increase of humidity, the growth mode changes as
A:{\it coexistence} $\rightarrow$
B:{\it uniform} $\rightarrow$
C:{\it periodic} $\rightarrow$ 
D:{\it branching}.
%
These changes result in production of the different domain patterns.
%
Figure \ref{fig:grad} shows change of the domain patterns 
formed in gradual increase of environmental humidity.
%
Here, thickness of the supersaturated solution film
can be characterized by the average area density $\rho$ [mg/cm$^{2}$].
%
$\rho$ is defined as the total amount of ascorbic acid in the film
divided by the film area.
%
These findings are common in both cases of methanol and aqueous solutions.
%
So far, we have obtained the two morphology diagrams,
which are (i) a function of humidity 
and the average area density $\rho$ \cite{Ito03},
and (ii) a function of humidity and temperature \cite{Yamazaki09}.
%
The diagram (ii) is shown in Fig. \ref{fig:diagram}.

%

%


Regarding the periodic growth, 
the following experimental results 
and theoretical explanations were previously reported.
%
In 1984, Iwamoto et al. presented a concentric ring pattern 
of ascorbic acid crystal domain from its methanol solution, 
which was formed in a refrigerator \cite{Iwamoto84}.
%
However in 1996, Fukunaga pointed out
that the concentric ring pattern found by Iwamoto et al. was an artifact, 
since the periodic growth synchronized with the periodic temperature regulation
in a refrigerator \cite{Fukunaga96}.
%
Instead, Fukunaga found an essentially different type of periodic growth,
now we are focusing on.
%
He also found various domain patterns, 
obtained in the uniform, periodic, and branching growth modes,
depending on humidity as described above.
%
In 2002, Uesaka et al. suggested that the two factors were important
for the periodic growth found by Fukunaga \cite{Uesaka02, Uesaka03}:
(1) a local fluidization of ``amorphous''-like solution
in the vicinity of the domain growth front in crystallization, and
(2) existence of fine spacings between the crystals forming the domain.
%
Based on these two factors,
they proposed the following scenario for the periodic growth.
%
The solution in the vicinity of the growth front became easy to flow by local fluidization, 
and it was absorbed into the fine spacings between the crystals
for their capillary force.
%
Then, the fluidized solution vanished in front of the domain 
and the crystal growth stopped.
%
However, the crystal growth was able to restart 
if there were some local points where the solution was in contact with the crystals.
%
Furthermore, they reported that the local fluidization did not occur 
in the uniform growth \cite{Uesaka02}.
%
They concluded that the difference
between the uniform and periodic growth modes was whether
the local fluidization occurred or not.


%
In this paper, we report our results of further investigation 
on the dynamical properties and humidity dependences 
of the domain front motion and solution flow in crystallization of ascorbic acid
with some reviews of the previous studies.
%
Especially, we focus on 
the uniform and periodic growth modes.
%
The rest of this paper is organized as follows.
%
In \S\ref{sec:exp}, our observation results are reported.
%
In order to grasp the essential features of the dynamical behaviors, 
we propose a simple toy model in \S\ref{sec:scenario}.
%
Some related topics are discussed in \S\ref{sec:discuss}.
%
Summary and conclusion are given in \S\ref{sec:conclusion}.

\section{Experiments}
\label{sec:exp}

\subsection{Setup}
\label{sec:setup}

L(+)-ascorbic acid (cica-reagent, Kanto Chemical, Japan)
was dissolved in purified water.
%
The solutions were spread uniformly
on a glass plate which was precleaned by Cica Clean LX-III (Kanto Chemical, Japan).
%
The amount of ascorbic acid is specified
by the average area density $\rho$ [mg/cm$^{2}$].
%
The glass plate with the solution was set horizontally
under a constant humidity $H$ [\%] and a constant temperature $T$ [$^{\circ}$C].
%
The crystal growth occurred due to water evaporation.
%
Measurements of $H$ and $T$ were done
by a hygrometer and a thermometer
in a data logger system (Easy sense II, Narika, Japan).
%
In our experiment, $T$ was fixed at a constant 
in the range from 25 $^{\circ}$C to 30 $^{\circ}$C.
%
From Fig. \ref{fig:diagram}, it is found 
that the value of $H$ at which the growth mode changes from uniform to periodic
is almost constant ($\approx 60\%$) within this temperature range.
%
The obtained domain patterns and their growth fronts 
were observed by using
a stereomicroscope (SZX-12, Olympus, Japan)
and a phase contrast microscope (CKX-41, Olympus, Japan).
%
The height profile of the crystal domain
was measured with a laser microscope (VK-9710, Keyence, Japan).
%
Domain pattern formation was recorded
by a time lapse recorded system with 30 frames per second 
(SVT-S960ES, Sony, Japan).

\subsection{Growth front motion}
\label{sec:motion}

\begin{figure}[!p]
\begin{center}
\includegraphics*[height=4.8cm]{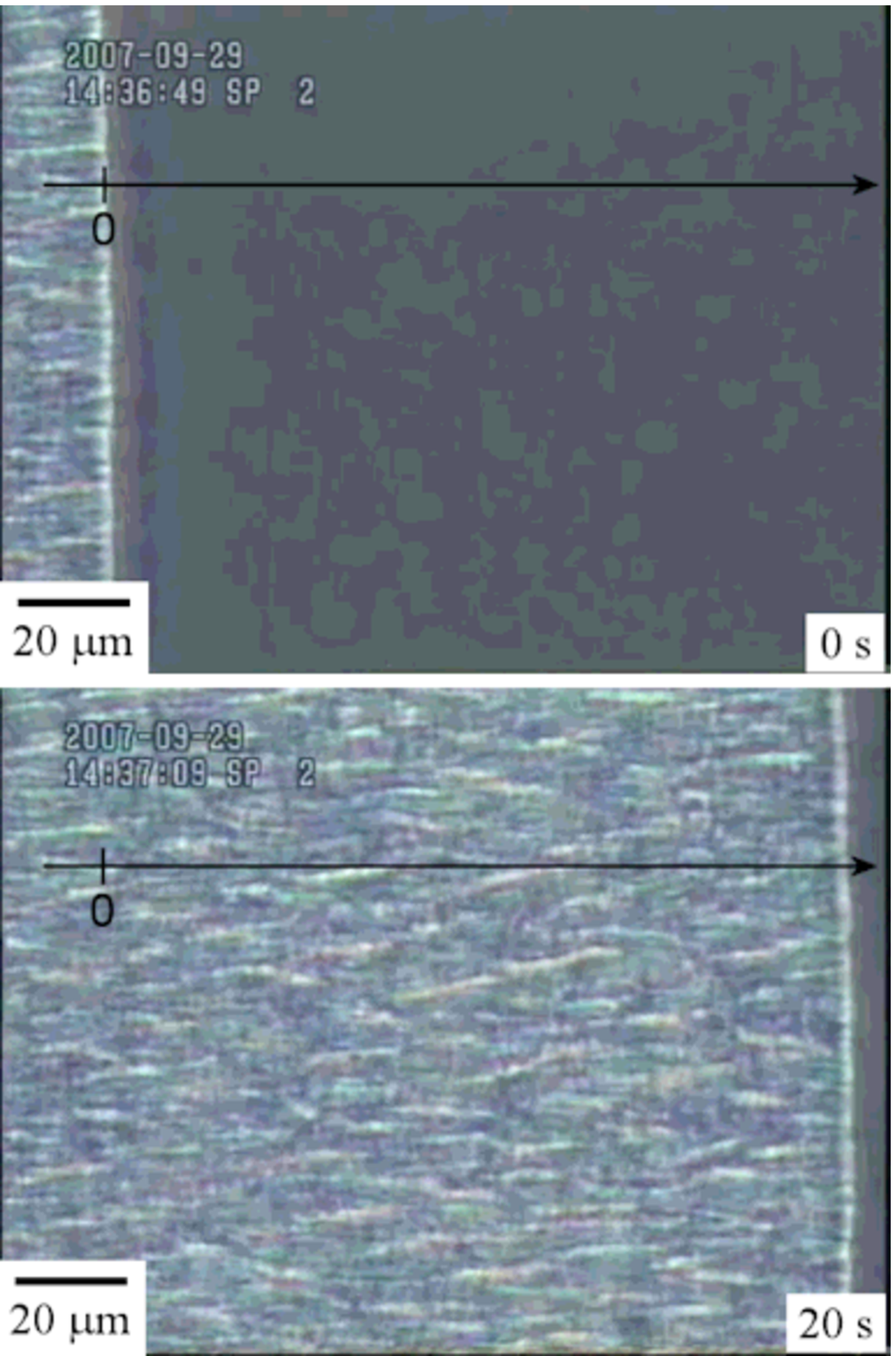}
\includegraphics*[height=4.8cm]{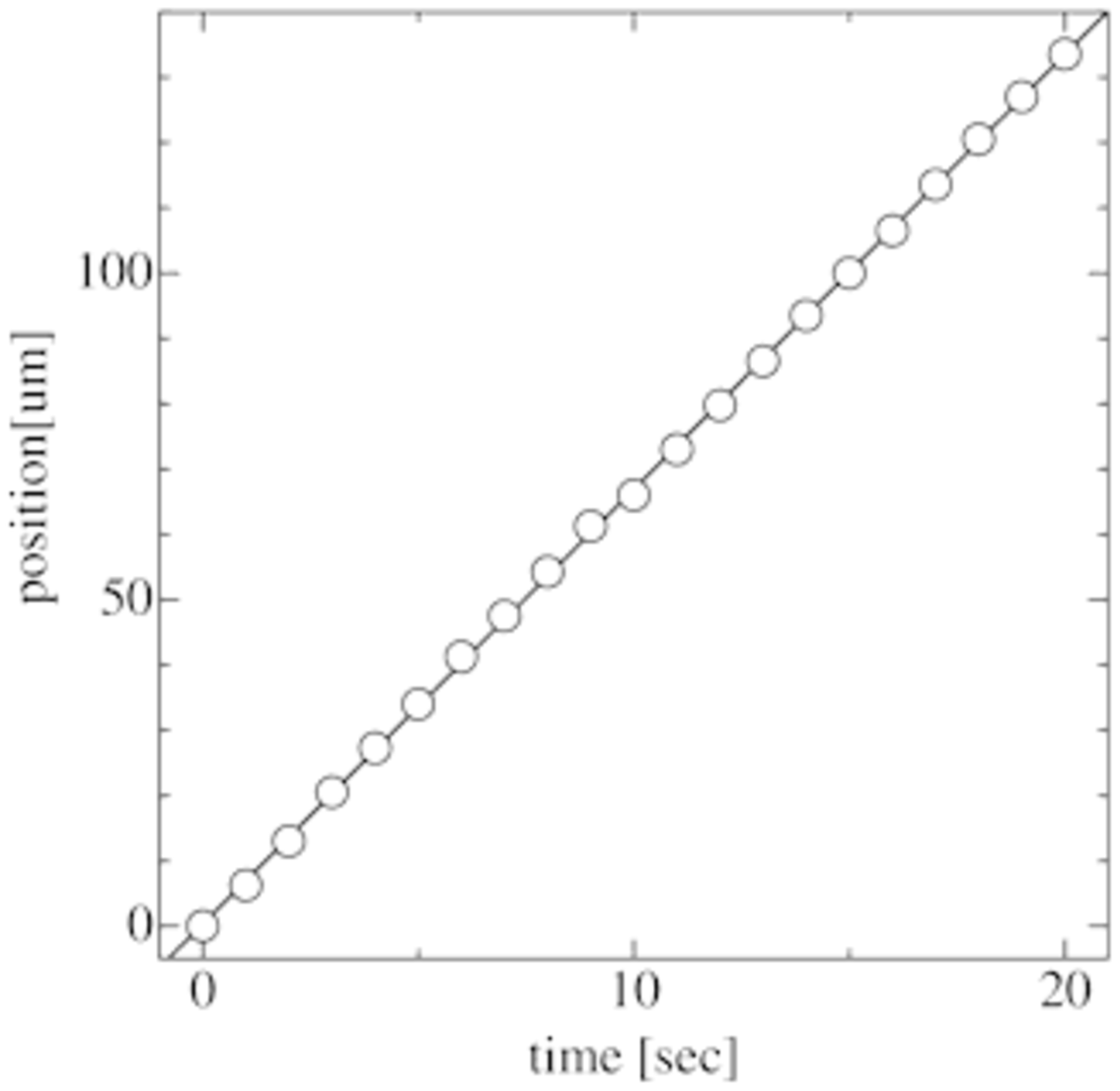}\\
(a) \hspace{5cm} (b)\\
\caption{Domain front motion in the uniform growth ($H =$ 50 \%, $\rho =$ 0.5 mg/cm$^{2}$, $T =$ 27 $^{\circ}$C).}
\label{fig:vel-u}
\end{center}
\end{figure}

\begin{figure}[!p]
\begin{center}
\includegraphics*[height=4.8cm]{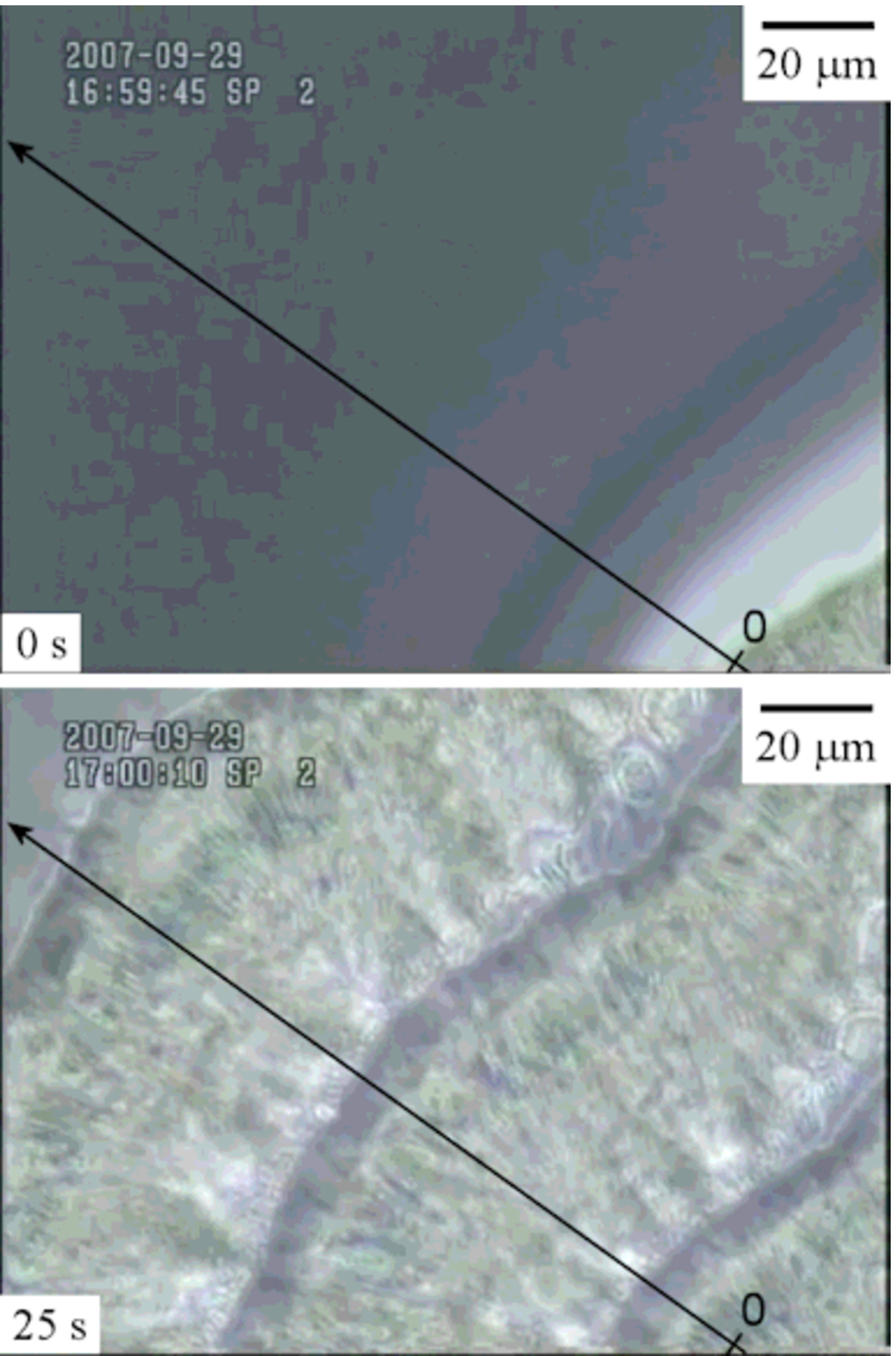}
\includegraphics*[height=4.8cm]{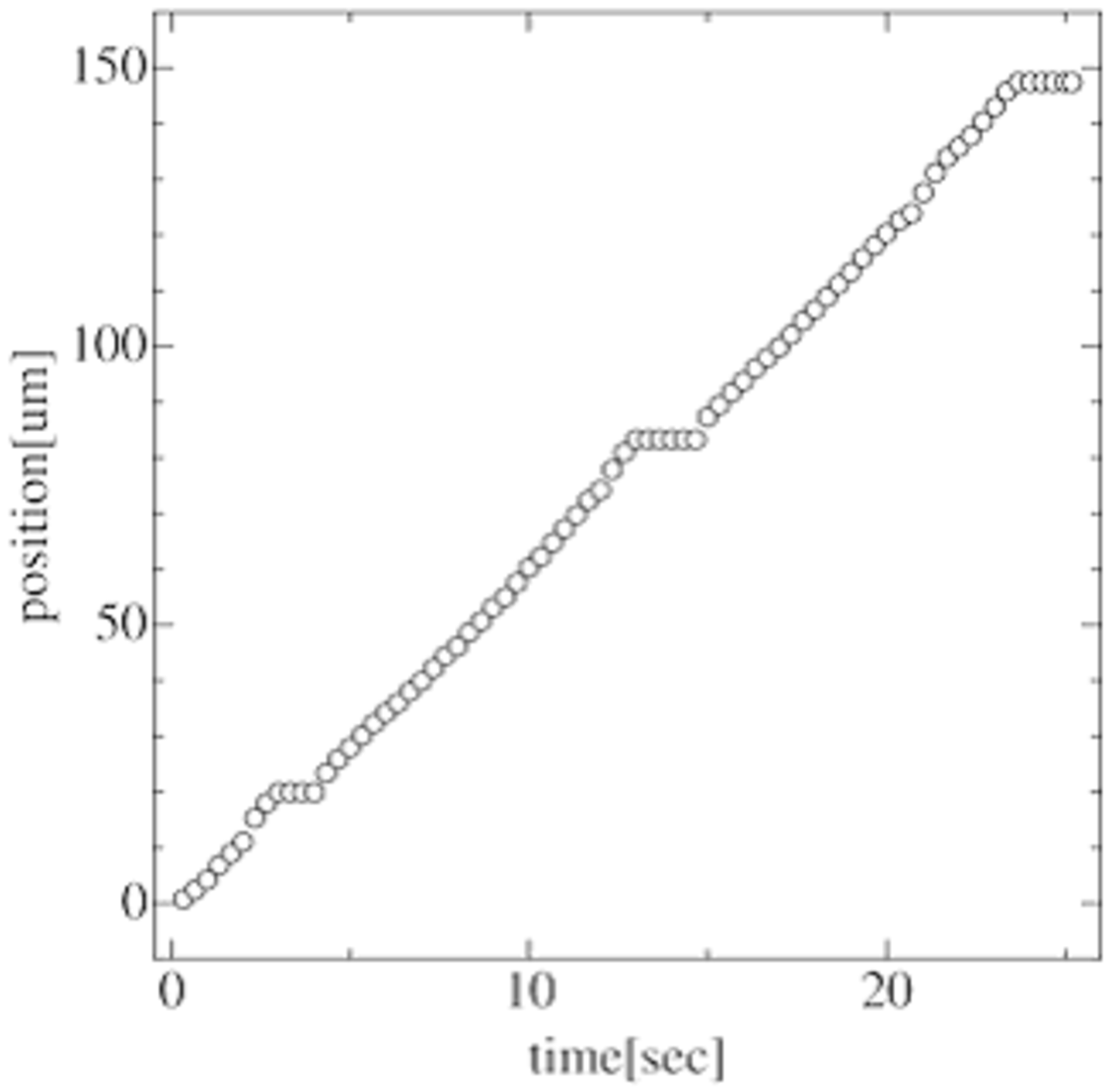}\\
(a) \hspace{5cm} (b)\\
\caption{Domain front motion in the periodic growth ($H =$ 65 \%, $\rho =$ 0.5 mg/cm$^{2}$, $T =$ 27 $^{\circ}$C).}
\label{fig:vel-p}
\end{center}
\end{figure}

\begin{figure}[!p]
\begin{center}
\includegraphics*[height=4.8cm]{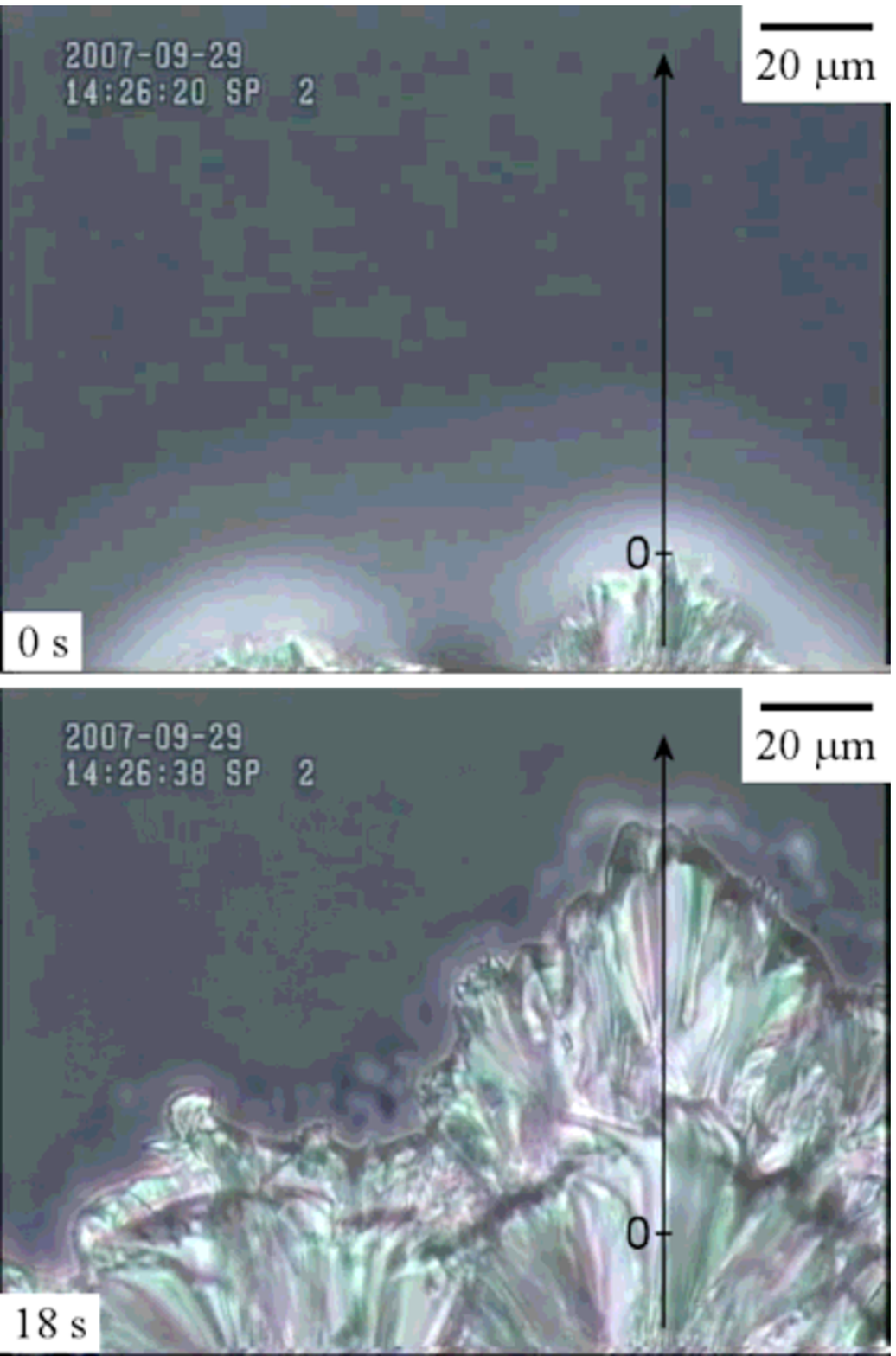}
\includegraphics*[height=4.8cm]{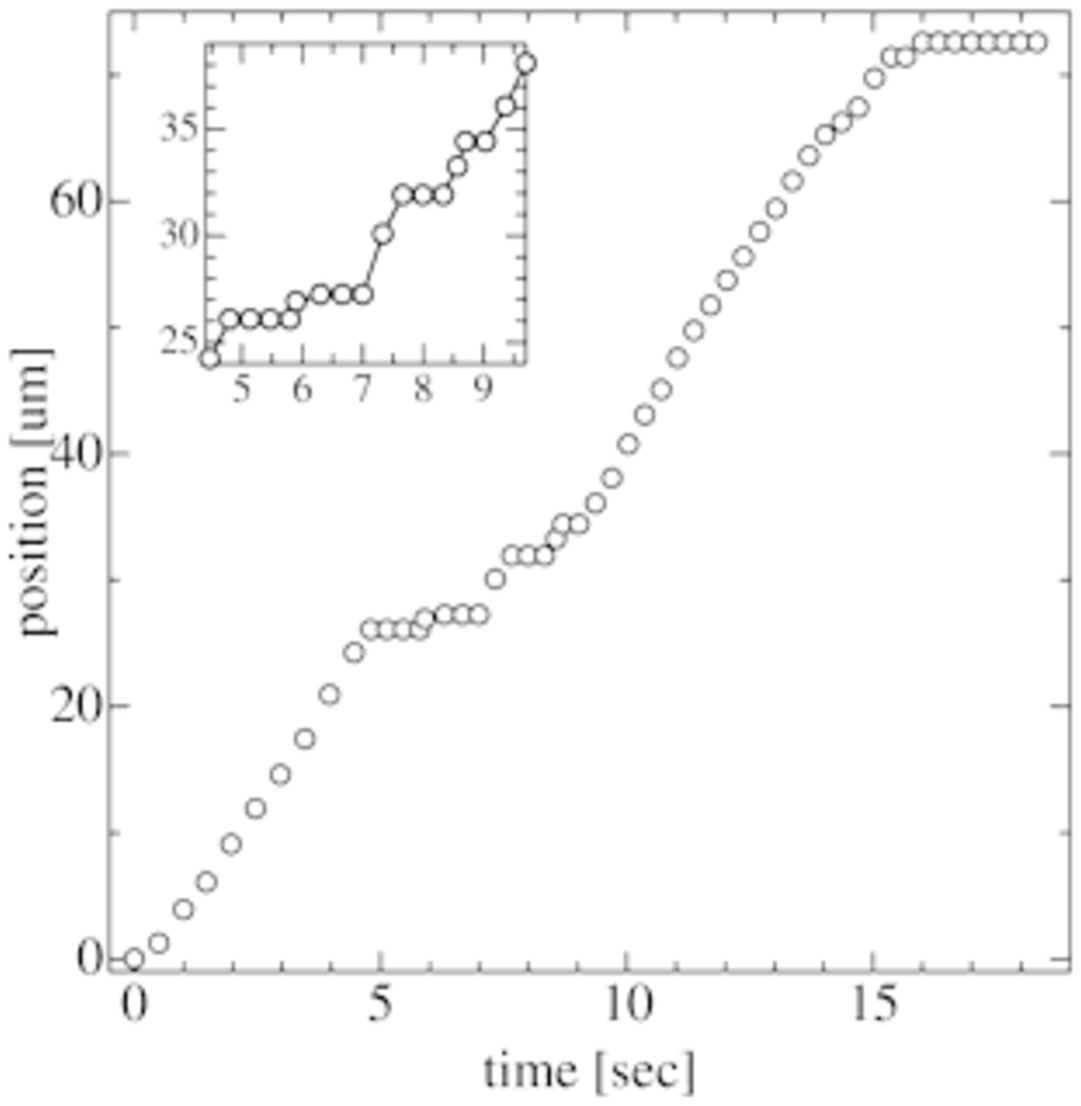}\\
(a) \hspace{5cm} (b)\\
\caption{Domain front motion in the branching growth ($H =$ 75 \%, $\rho =$ 0.5 mg/cm$^{2}$, $T =$ 27 $^{\circ}$C).}
\label{fig:vel-b}
\end{center}
\end{figure}

%
Figures \ref{fig:vel-u}-\ref{fig:vel-b} show (a) snapshots 
of a domain growth front and (b) its position on the axis as a function of time
in the uniform, periodic, and branching modes, respectively.
%
In these figures, $\rho$ and $T$ are fixed.
%
The positive linear slopes in the panel (b) of each figure 
indicate that the fronts move at a constant speed.
%
From these figures, the values of speed in the uniform, periodic, and branching 
modes are obtained as about 6.7 $\mu$m/s, 6.8 $\mu$m/s, and 6.3 $\mu$m/s, 
respectively.
%
This result suggests that if $\rho$ and $T$ are fixed 
then the front speed is almost constant when 50 \% $\lesssim H \lesssim$ 75 \%,
although the growth mode changes.
%
As shown in Fig. \ref{fig:vel-p}, the periodic mode consists of 
the advance and suspension periods alternating regularly.
%
This result is consistent with the previous reports\cite{Fukunaga96, Uesaka02}.
%
It is also found from Fig. \ref{fig:vel-b} that the branching mode
also possesses the advance and suspension periods.
%
However, unlike the periodic mode, occurrence of the two periods becomes irregular 
(see the inlet of Fig. \ref{fig:vel-b} (b)).
%
The frequency of such irregularity increased as $H$ rose.

\begin{figure}[!p]
\begin{center}
\includegraphics*[height=4.8cm]{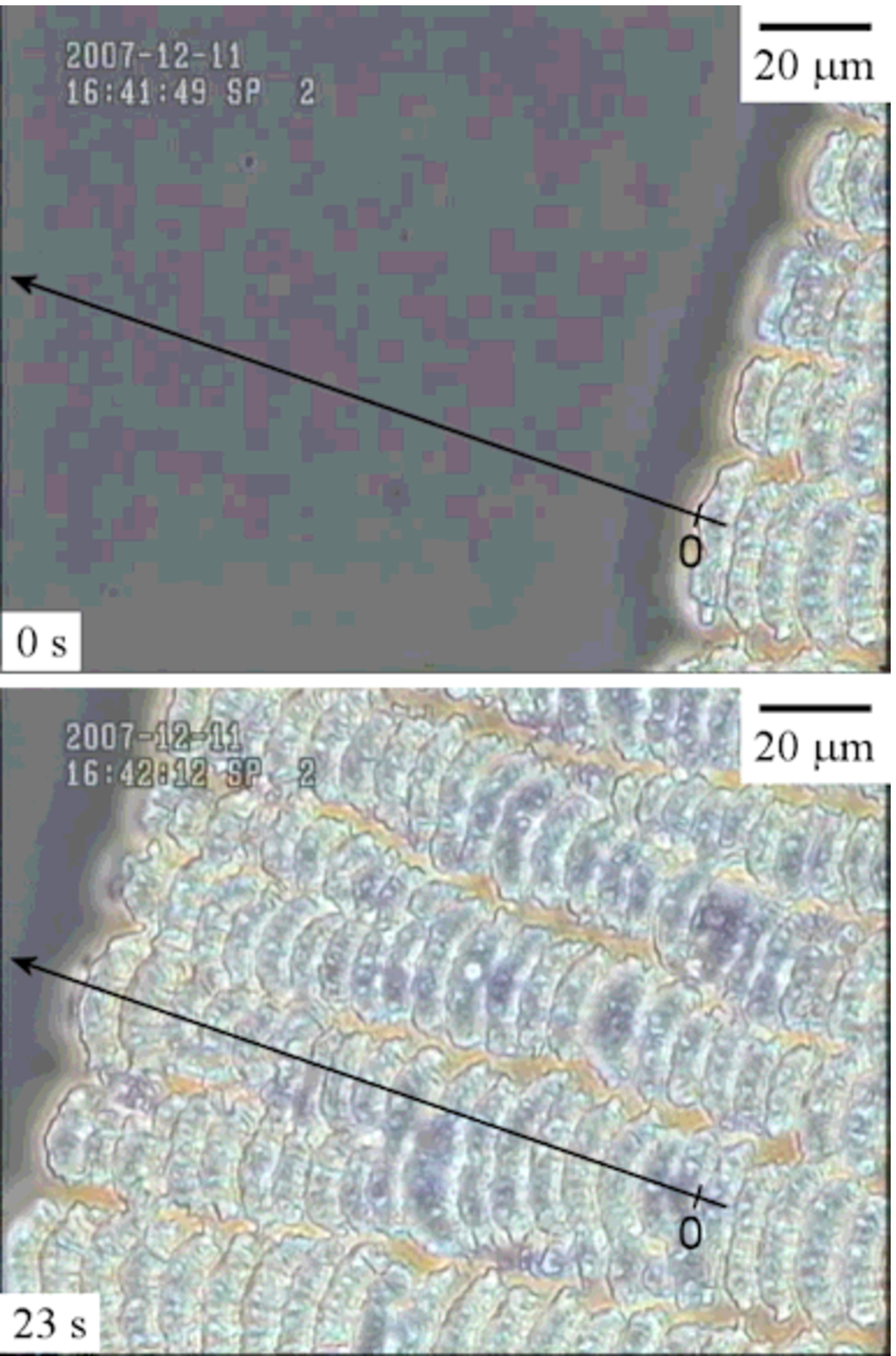}
\includegraphics*[height=4.8cm]{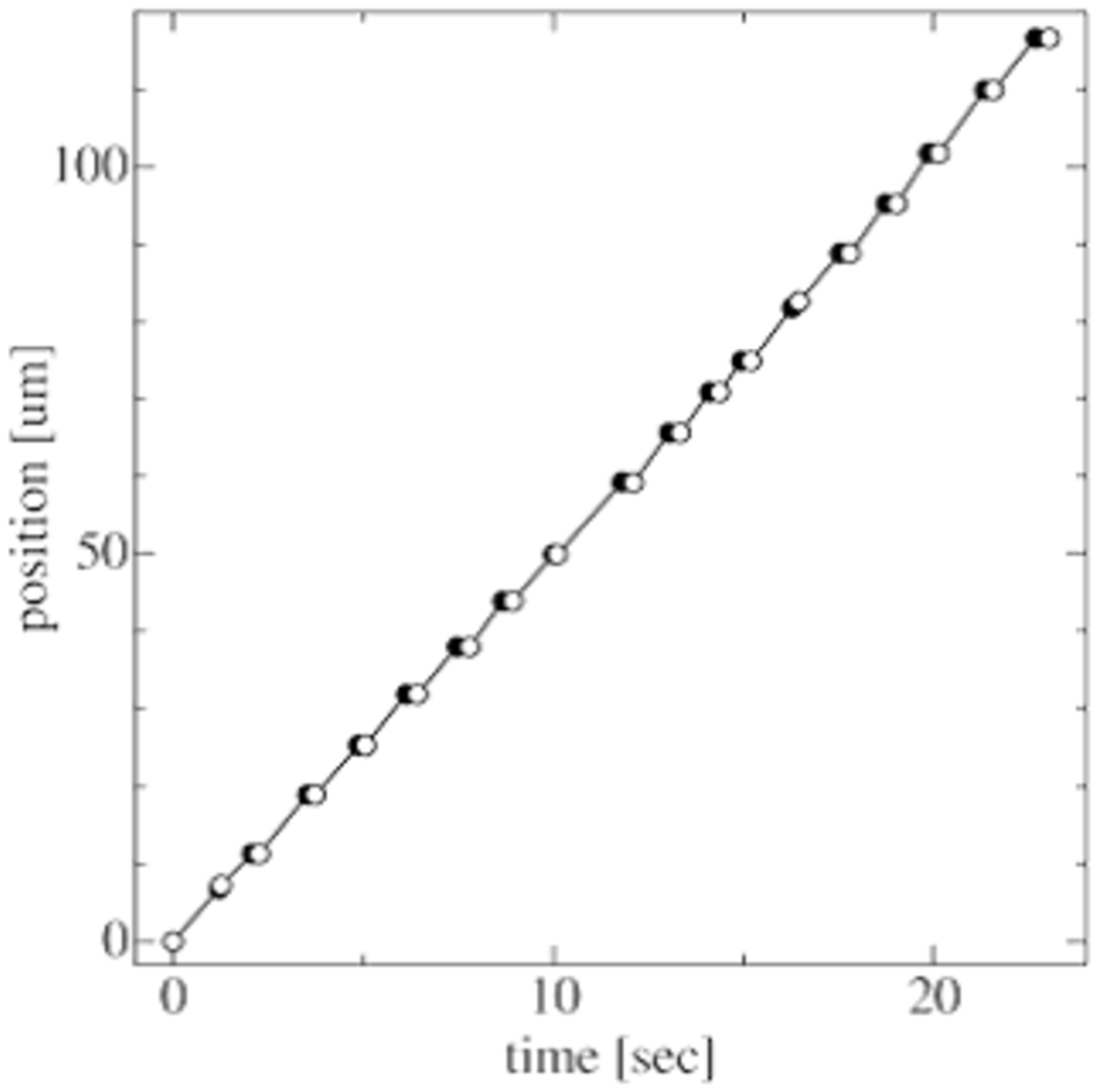}\\
(a) \hspace{5cm} (b)\\
\caption{Domain front motion in the periodic growth ($H =$ 65 \%, $\rho =$ 0.2 mg/cm$^{2}$, $T =$ 25 $^{\circ}$C).
The filled and open circles show the beginning of the suspension and the advance periods, 
respectively.}
\label{fig:vel-p-th}
\end{center}
\end{figure}

\begin{figure}[!p]
\begin{center}
\includegraphics*[height=4.8cm]{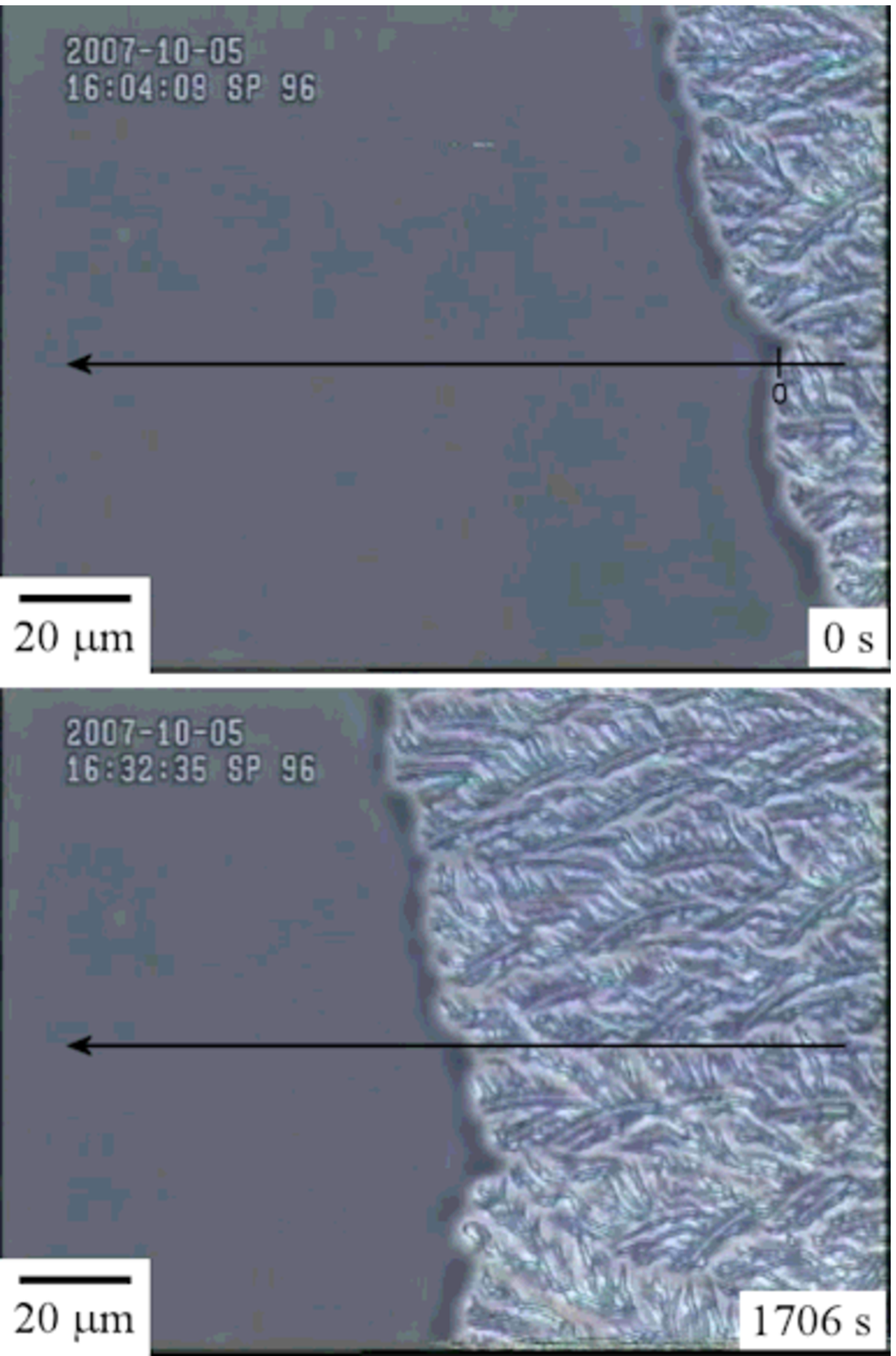}
\includegraphics*[height=4.8cm]{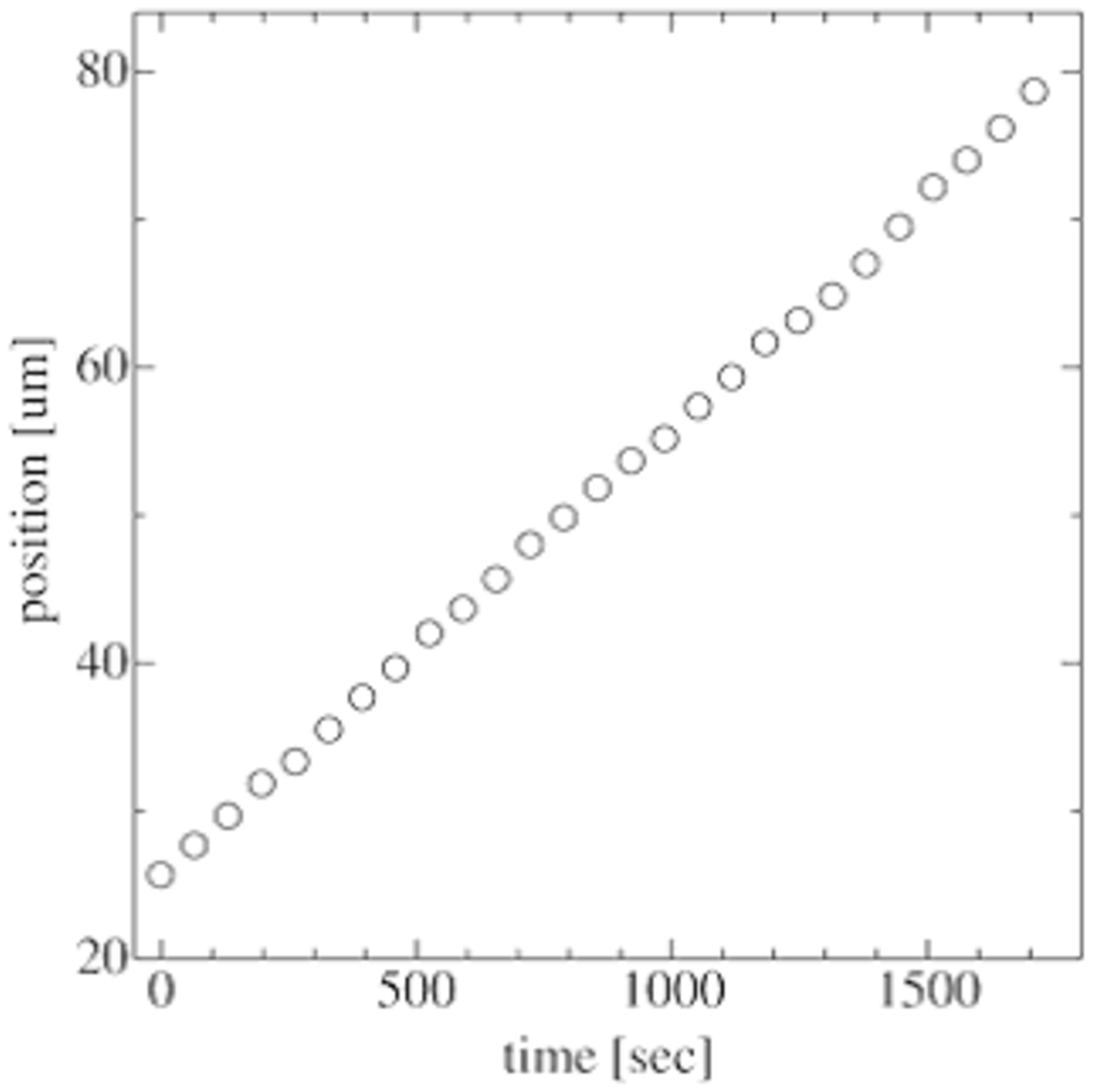}\\
(a) \hspace{5cm} (b)\\
\caption{Dense branching morphology formation
 ($H =$ 35 \%, $\rho$ is less than 0.1 mg/cm$^{2}$, $T =$ 27 $^{\circ}$C)}.
\label{fig:vel-u-b}
\end{center}
\end{figure}

%
In our observation, with decrease of $\rho$, 
the pitch of concentric rings became narrower 
and the suspension period became shorter 
(see Fig. \ref{fig:vel-p-th} for reference).
%
From this figure, the front speed in the advance period
is estimated as about 5.1 $\mu$m/s, which is slower 
than those in Figs. \ref{fig:vel-u}-\ref{fig:vel-b}.
%
Figure \ref{fig:vel-u-b} shows a case 
where $\rho$ is less than 0.1 mg/cm$^{2}$.
%
The domain front speed is found to be constant and is estimated 
as 3.0 $\times$ 10$^{-2}$ $\mu$m/s,
which is about 1/100 of that in Fig. \ref{fig:vel-u}.
%
Note that the condition of $H$ and $T$ corresponds to the uniform growth mode
as shown in Fig. \ref{fig:diagram}, although $\rho$ is different.
%
In spite of that, Fig. \ref{fig:vel-u-b} shows that 
the obtained domain exhibits a dense branching pattern.

%

%

%

%

%

\subsection{Height profile in the periodic growth}
\label{sec:height}

\begin{figure}[!p]
\begin{center}
\includegraphics*[width=4cm]{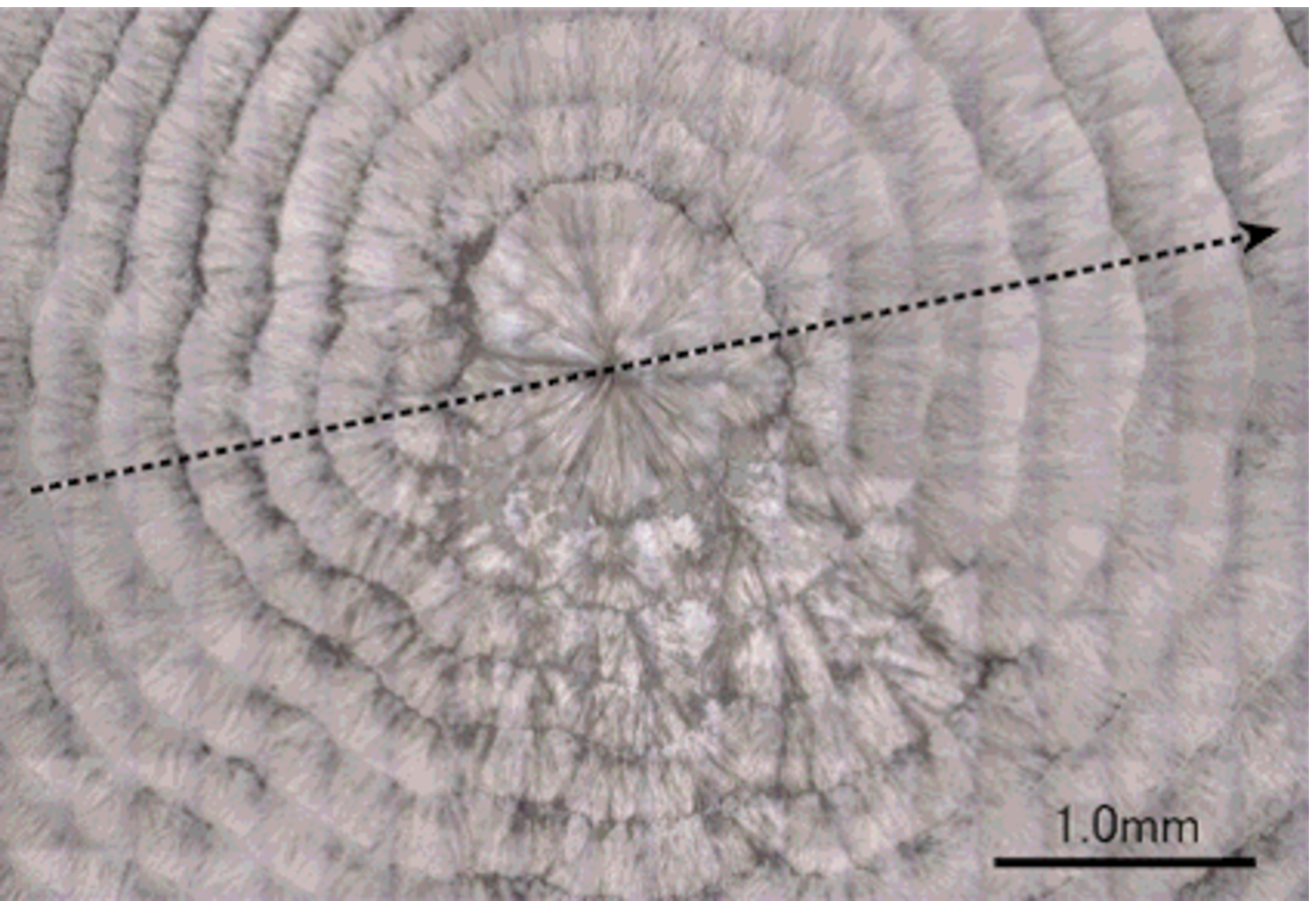}
\includegraphics*[width=4cm]{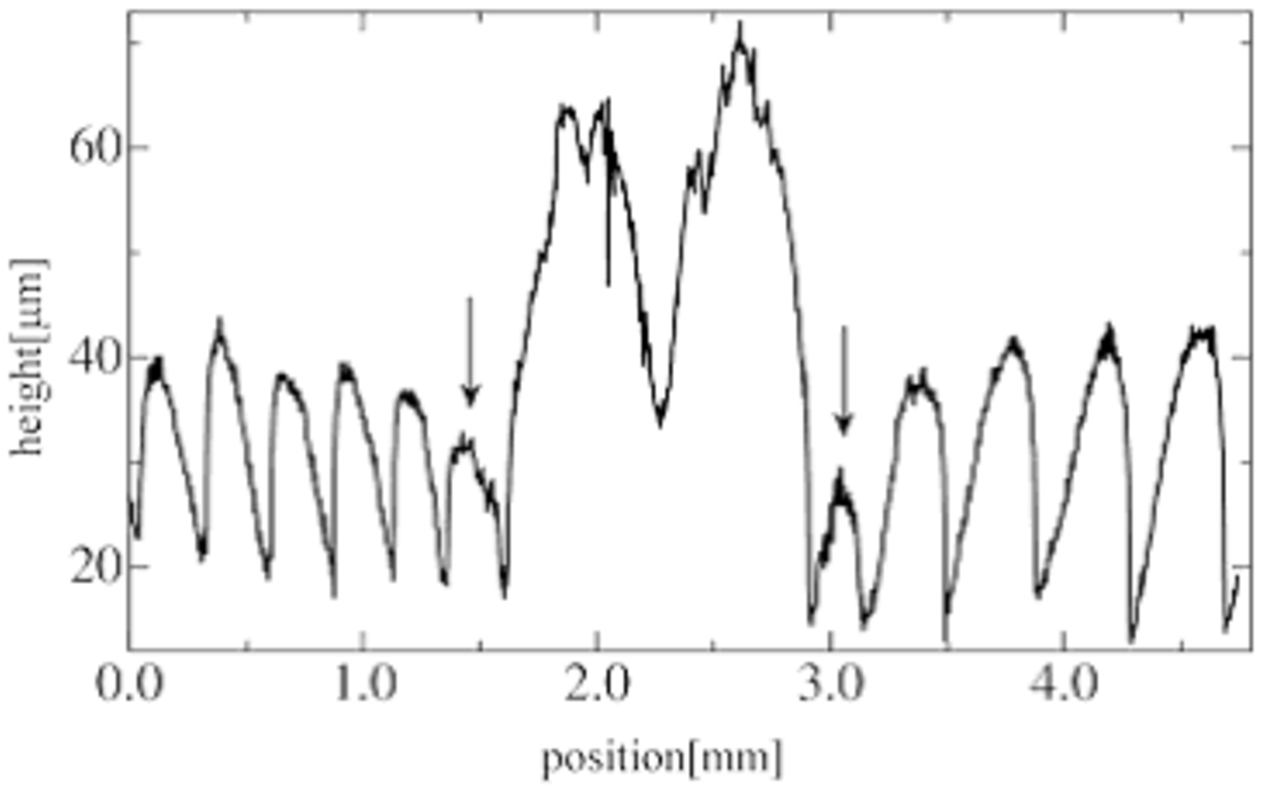}\\
(a) \hspace{3cm} (b)
\caption{(a) A typical concentric pattern by the periodic growth 
($H =$ 65 \%, $\rho =$ 1.5 mg/cm$^{2}$, $T =$ 30$^{\circ}$C). 
(b) The height profile along the dotted line in (a).} 
\label{fig:h-p}
\end{center}
\end{figure}

Here we show the height profile of concentric ring patterns 
obtained by the periodic growth.
%
A typical concentric ring pattern is shown in Fig. \ref{fig:h-p}(a).
%
The height profile measured along the dotted arrow is plotted
in the panel (b).
%
The following features are confirmed.
%
(i) The height periodically varies 
in accordance with the pitch of the concentric rings.
%
(ii) The peak heights are almost constant except for the central region
and for the first ring adjacent to the central.
(The first ring is indicated by the arrows in the panel (b).)
%
(iii)
The height of the central region becomes higher.
%
On the other hand, the peak of the first ring becomes lower.

\begin{figure}[!p]
\begin{center}
\includegraphics*[width=6cm]{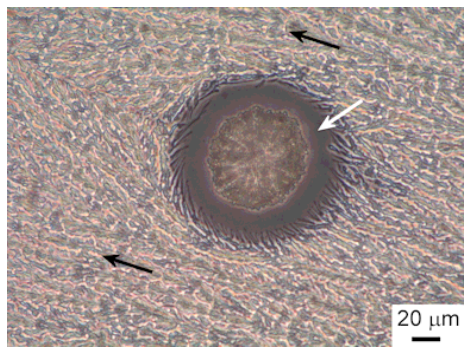}
\caption{The dark ring region pointed by the white arrow shows no crystals. 
The black arrows show the direction of crystallization after the formation of the ring region
($H =$ 60 \%, $\rho =$ 0.5 mg/cm$^{2}$, $T =$ 27$^{\circ}$C). 
}
\label{fig:eye}
\end{center}
\end{figure}

%
These observation results can be explained as follows.
%
Crystallization brings about solution flow directed to the crystal domain.
%
And due to supply of the solute by the solution flow, 
the height of the central region increases.
%
On the other hand, 
thickness of the solution around the central region becomes thin 
for the solution flow.
%
Then, the peak height of the first ring becomes low.
%
In an extreme case, 
the circular domain is isolated from the solution and stops to grow
as shown in  Fig. \ref{fig:eye}; 
%
The ring region without the solution is formed by 
the solution flow in crystallization.

%

%

\subsection{Solution flow in crystallization: existence of threshold}
\label{sec:flow}

\begin{figure}[!p]
\begin{center}
\includegraphics*[height=3.5cm]{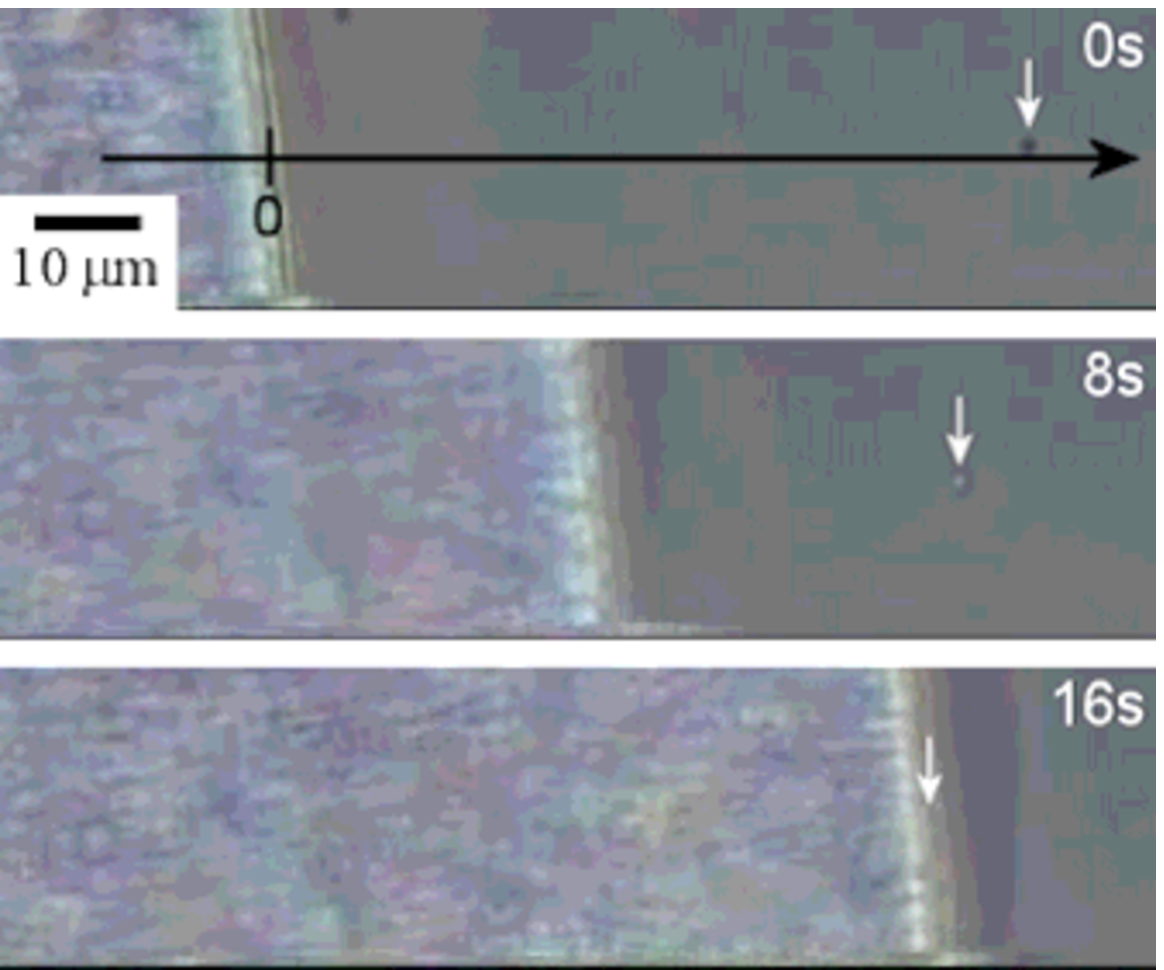}
\includegraphics*[height=3.5cm]{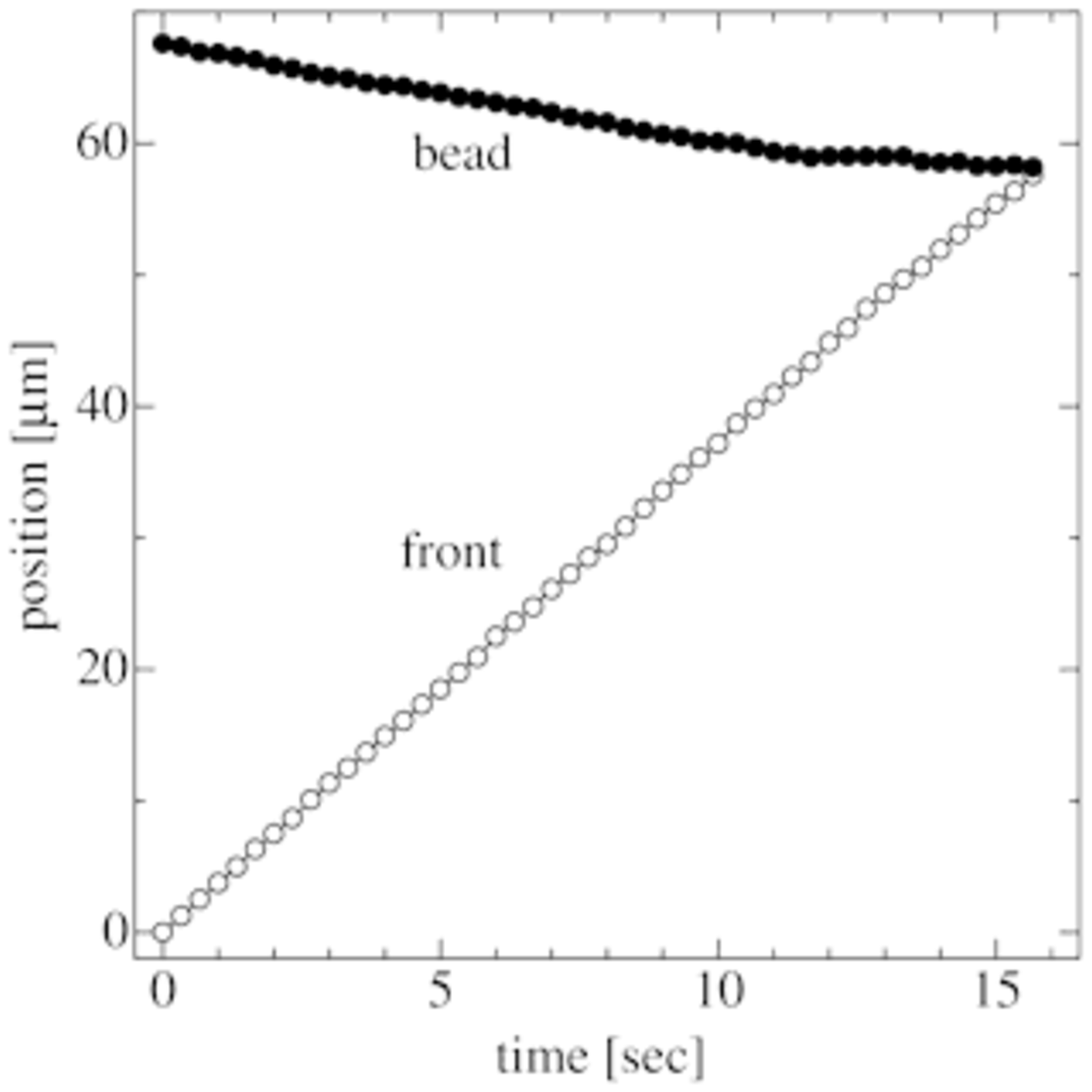}\\
(a) \hspace{3cm} (b)
\caption{Bead and domain front motion in the uniform growth 
($H =$ 30 \%, $\rho =$ 0.5 mg/cm$^{2}$, $T =$ 25 $^{\circ}$C).
The white arrows in (a) point at the bead position.}
\label{fig:bd-u30}
\end{center}
\end{figure}

%
In order to visualize solution flow in crystallization,
we prepared aqueous solution of ascorbic acid with latex beads of 1$\mu$m diameter.
%
The motion of a bead showing the solution flow 
was observed with the phase contrast microscope described in \S\ref{sec:setup}.
%
%
Figure \ref{fig:bd-u30} shows the case of the uniform growth mode.
%
Positions of the growth front and the bead are plotted 
in the panel (b) as functions of time.
%
The bead motion shown in this figure
suggests existence of the solution flow directed to the domain, 
which occurs even about 60 $\mu$m distant from the domain front.
%
We confirmed that the flow 
had a tendency to become fast as $H$ rose.
%
%
%
%

\begin{figure}[!p]
\begin{center}
\includegraphics*[height=5.5cm]{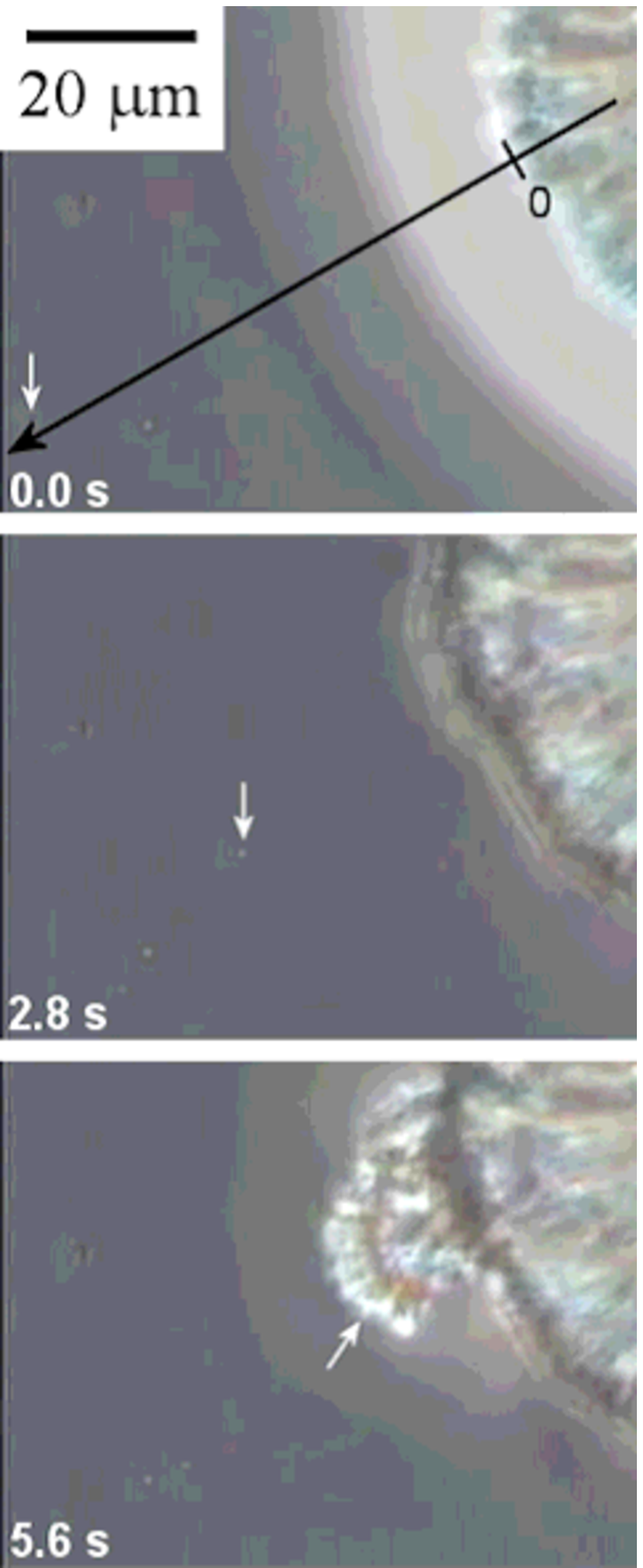}
\includegraphics*[height=5.5cm]{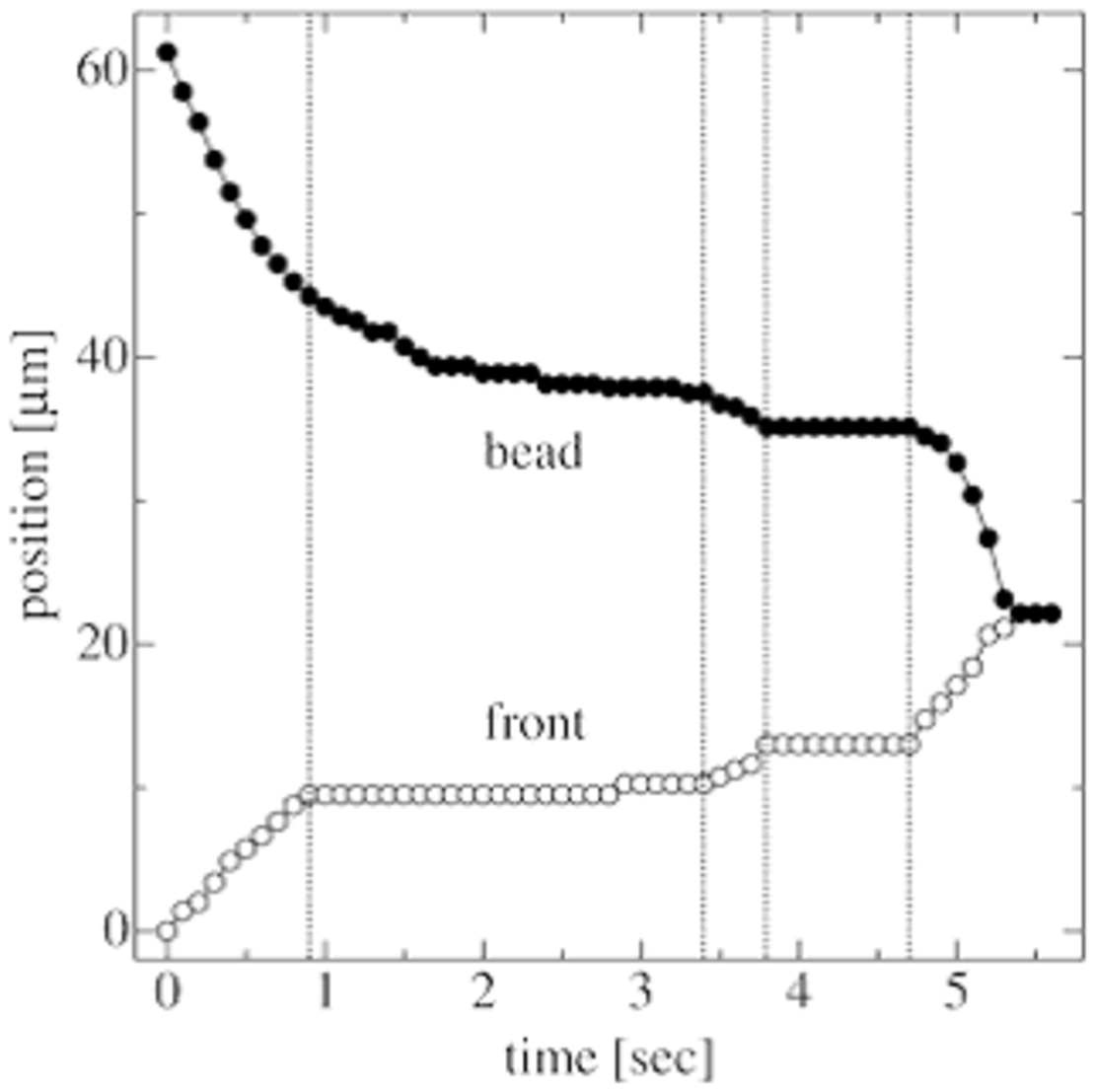}\\
(a) \hspace{3cm} (b)
\caption{Bead and domain front motion in the periodic growth 
($H =$ 65 \%, $\rho =$ 0.5 mg/cm$^{2}$, $T =$ 25 $^{\circ}$C).
The white arrows in (a) point at the bead position.}
\label{fig:bd-p}
\end{center}
\end{figure}

%
The case of the periodic growth mode 
is shown in Fig. \ref{fig:bd-p}.
%
From the panel (b), 
the following features can be confirmed for the bead motion and the domain front motion.
%
Once the growth front stops (at 0.9 sec in the panel (b)),
the bead motion also gradually stops.
%
And when the front moves again (at 4.7 sec), 
the bead also moves.
%
From this fact, it is concluded that
the solution flow is induced by the crystallization.

\begin{figure}[!p]
\begin{center}
\includegraphics*[height=4.5cm]{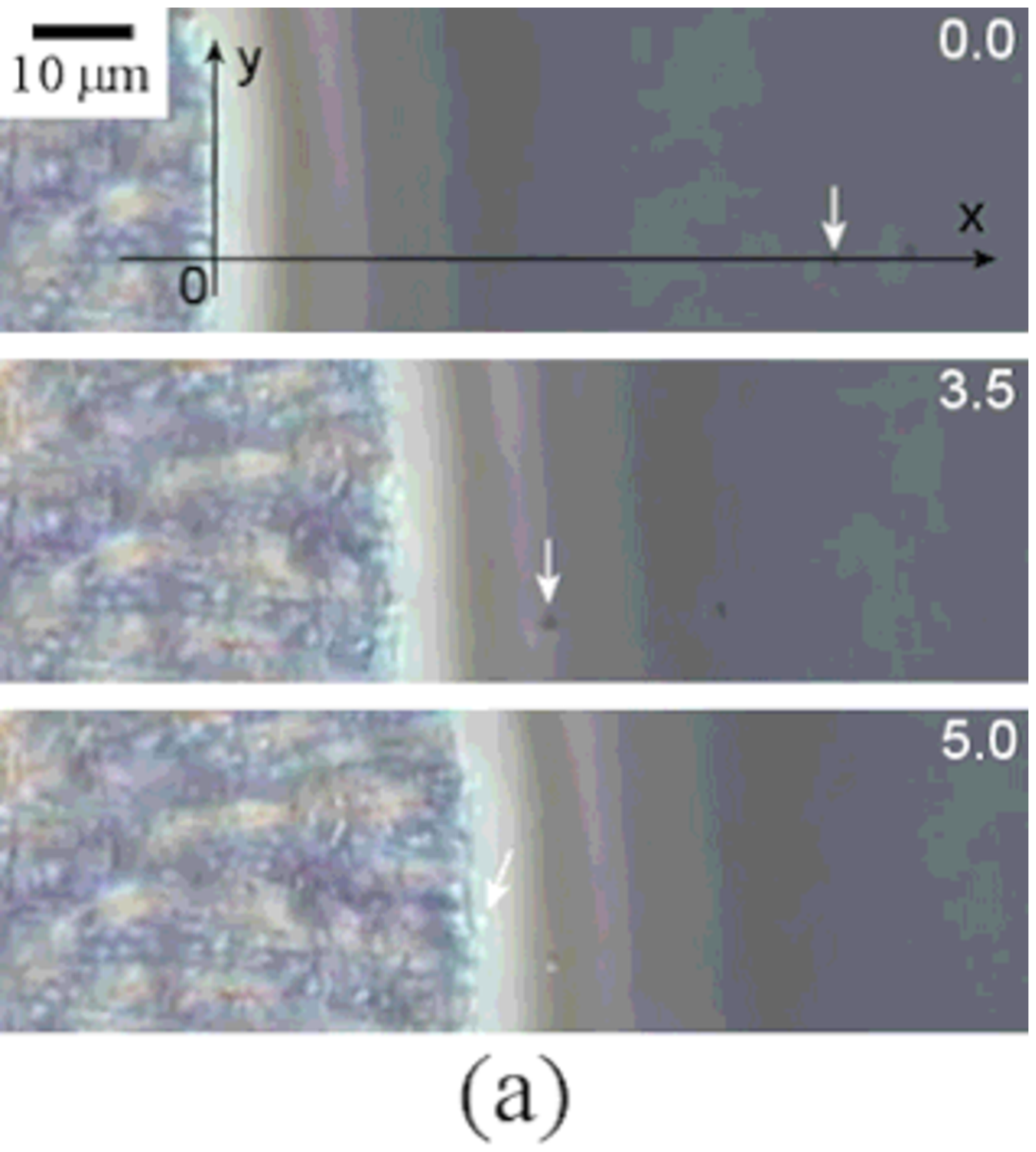}
\includegraphics*[height=4.5cm]{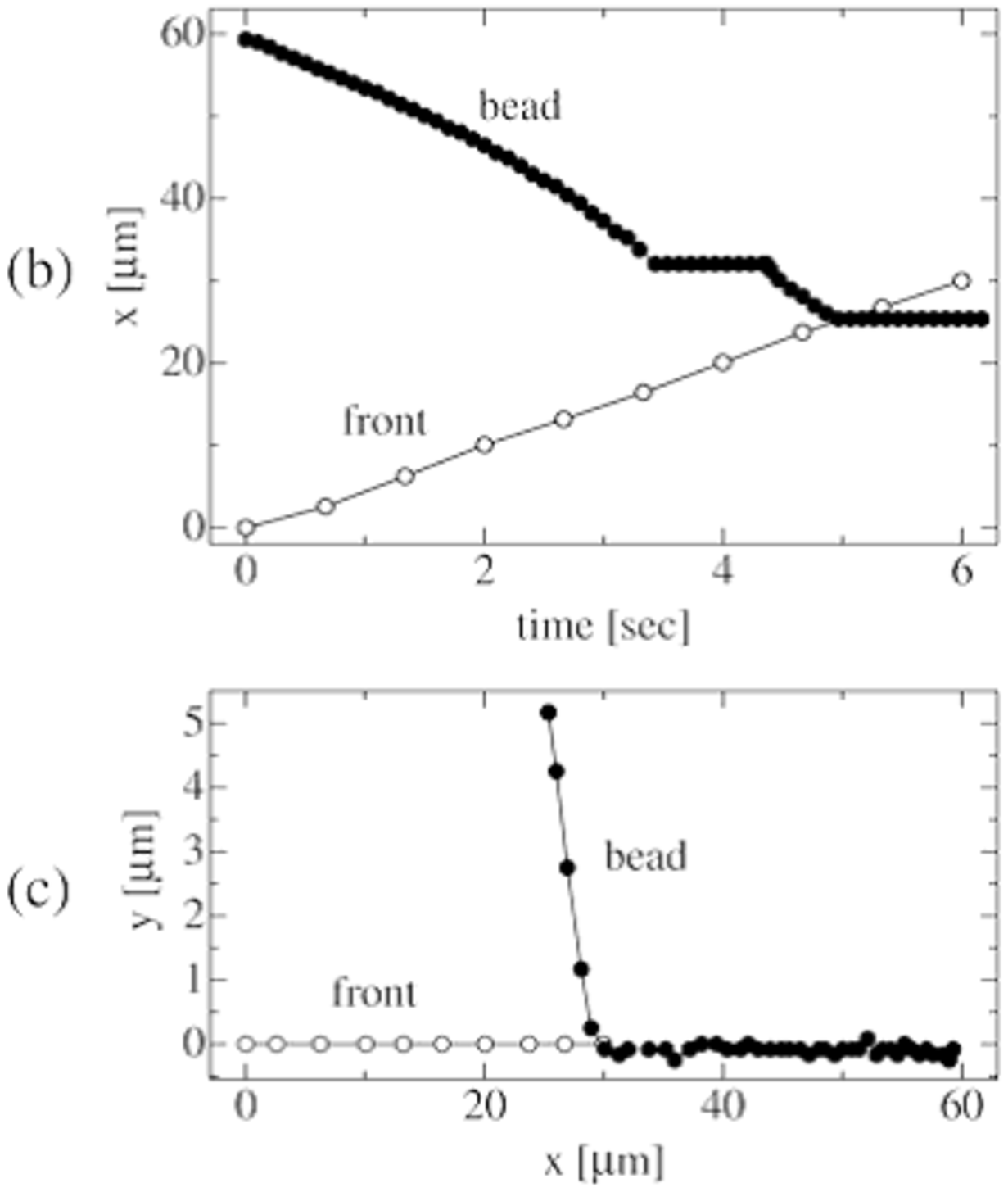}\\
\caption{Threshold-sensitive bead motion in the uniform growth 
($H =$ 50 \%, $\rho =$ 0.5 mg/cm$^{2}$, $T =$ 25 $^{\circ}$C).
The white arrows in (a) point at the bead position.}
\label{fig:bd-u50fl}
\end{center}
\end{figure}
\begin{figure}[!ph]
\begin{center}
\includegraphics*[height=6.5cm]{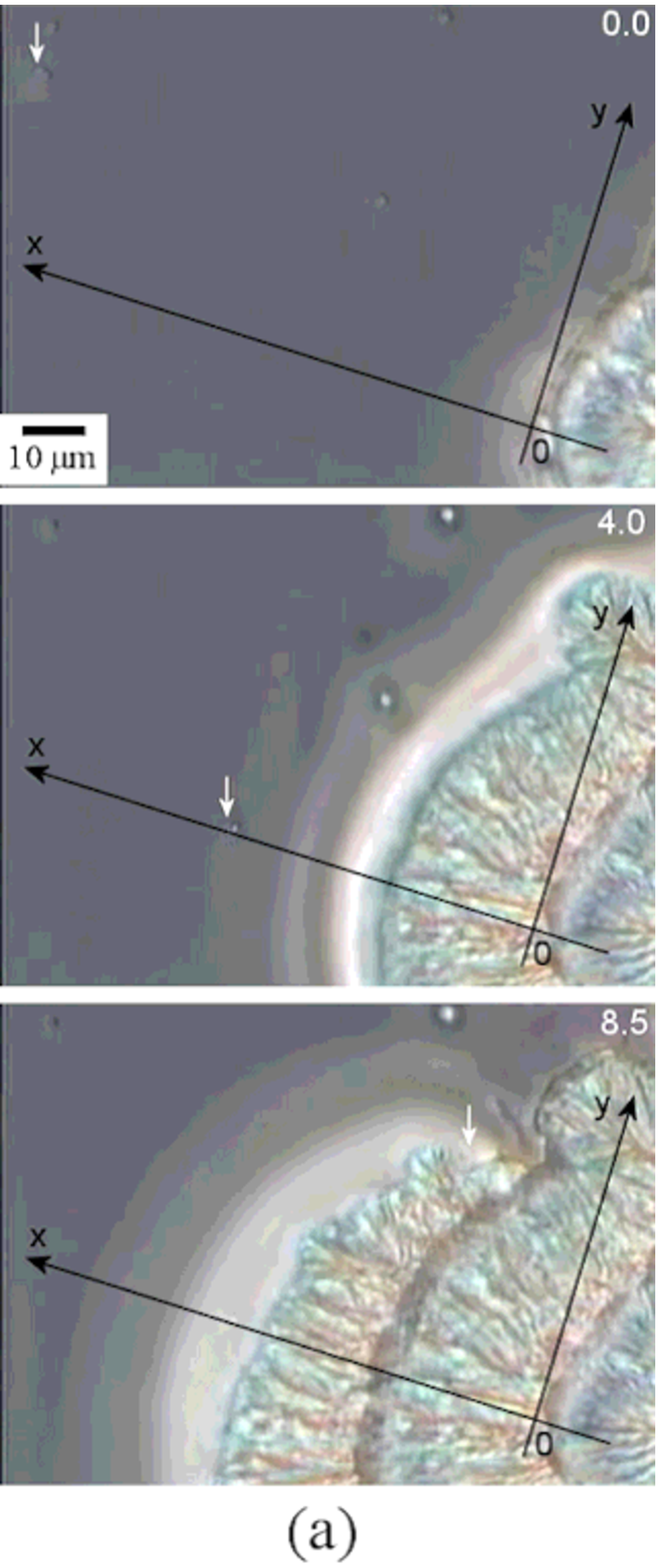}
\includegraphics*[height=6.5cm]{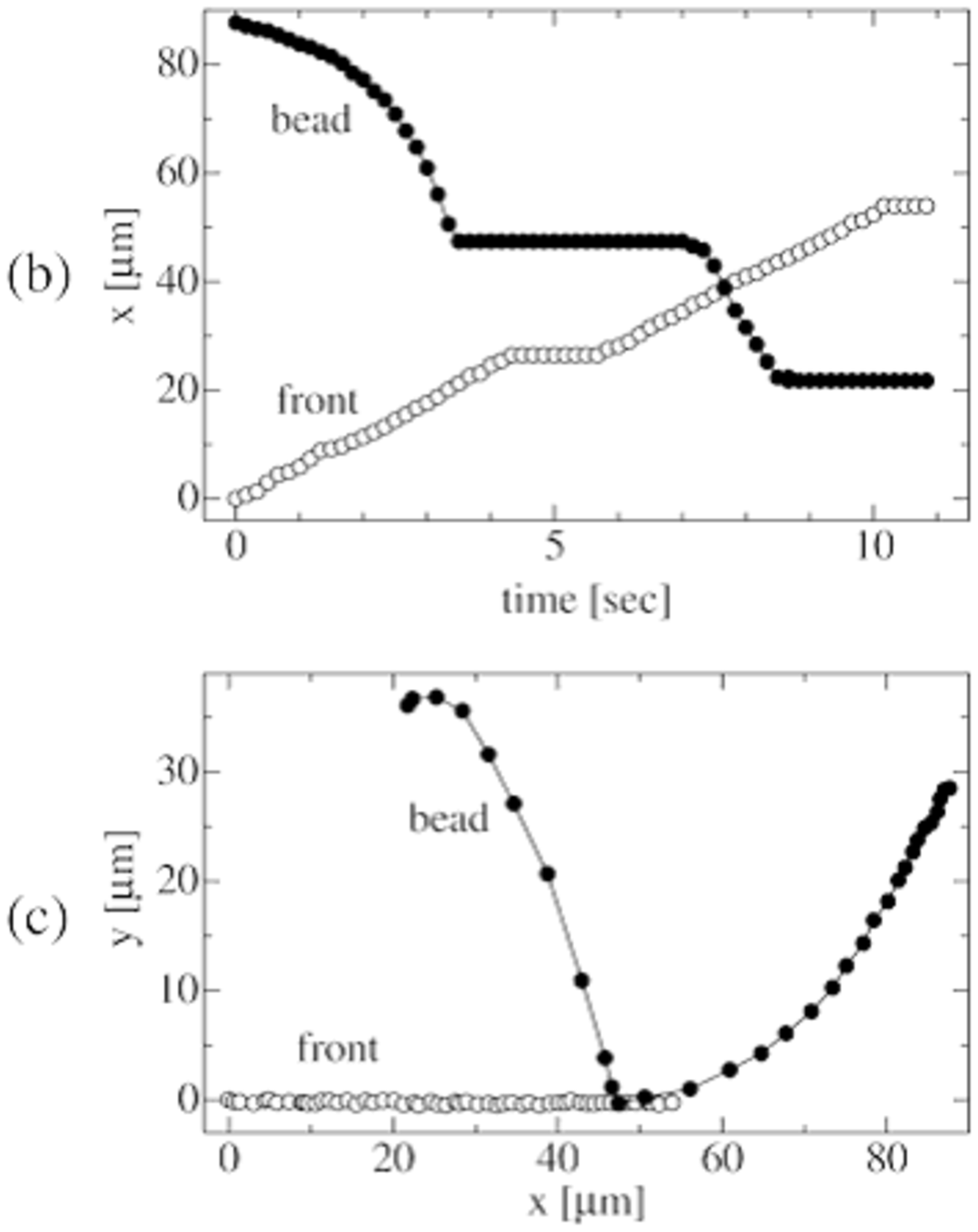}\\
\caption{Threshold-sensitive bead motion in the periodic growth 
($H =$ 65 \%, $\rho =$ 0.5 mg/cm$^{2}$, $T =$ 25 $^{\circ}$C).
The white arrows in (a) point at the bead position.}
\label{fig:bd-pfl}
\end{center}
\end{figure}

%
As a remarkable feature shown in Figs. \ref{fig:bd-u50fl} and \ref{fig:bd-pfl},
the ``stop-and-go'' motion of a bead was frequently observed.
%
When the bead and the growth front approach each other
and their distance becomes about 30 $\mu$m in this case,
the bead motion stops.
%
But the bead moves again when it is close to the front 
at about 10 $\mu$m distance.
%
This motion can be observed
in both the uniform mode (Fig. \ref{fig:bd-u50fl}) 
with a high humidity ($\gtrsim$ 50 \%)
and the periodic mode (Fig. \ref{fig:bd-pfl}).
%
Note that for the uniform growth mode, 
this motion does not occur in a low humidity case 
as shown in Fig. \ref{fig:bd-u30}.
%
Then, it is considered that 
the solution flow becomes strong and the solution film
in the vicinity of the domain front becomes thinner
as humidity becomes higher.
%
Furthermore, we confirmed that the occurrence frequency of 
this ``stop-and-go'' motion became high as $\rho$ decreased.
%
It is suggested from this result 
that the thickness of the solution film becomes thinner 
in the region between 10 and 30 $\mu$m distant from the growth front.
%
By considering the result shown in Fig. \ref{fig:vel-u-b} in addition,
existence of the threshold thickness for domain growth is suggested;
%
Below the threshold, supply of the solution to the domain growth front almost stops.
%
Furthermore, it is found from the panel (c) of Figs. \ref{fig:bd-u50fl} and \ref{fig:bd-pfl} 
that the bead within 10 $\mu$m distant from the growth front moves faster
and irregularly so that the motion is not always directed to the front,
but there are some cases where the bead moves along the periphery of the domain front.
%
This fact suggests that the fluidity of the solution
within 10 $\mu$m from the domain front becomes high.
%
Existence of this high fluidity region 
is consistent with the local fluidization 
reported by Uesaka et al.\cite{Uesaka02, Uesaka03}.
%
Note that this ``stop-and-go'' motion of the bead 
is caused by the thickness threshold and 
is essentially different from the bead motion
synchronized with the periodic growth shown in Fig. \ref{fig:bd-p}.

%

%

%

%

%

\subsection{Collision of domain fronts: self-similarity, synchronization, 
and threshold sensitivity}

\begin{figure}[!p]
\begin{center}
\includegraphics*[width=5cm]{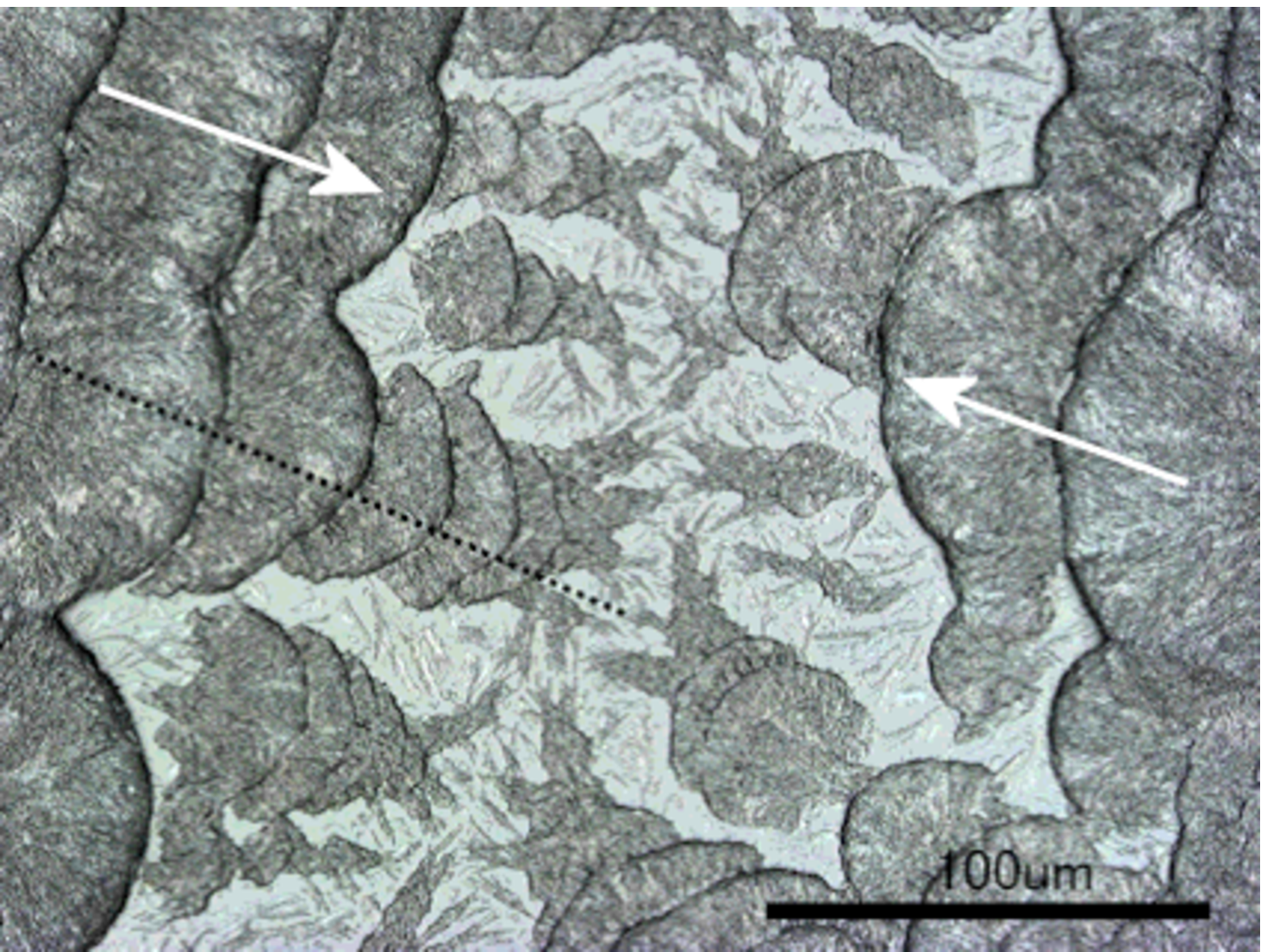}
\includegraphics*[width=5cm]{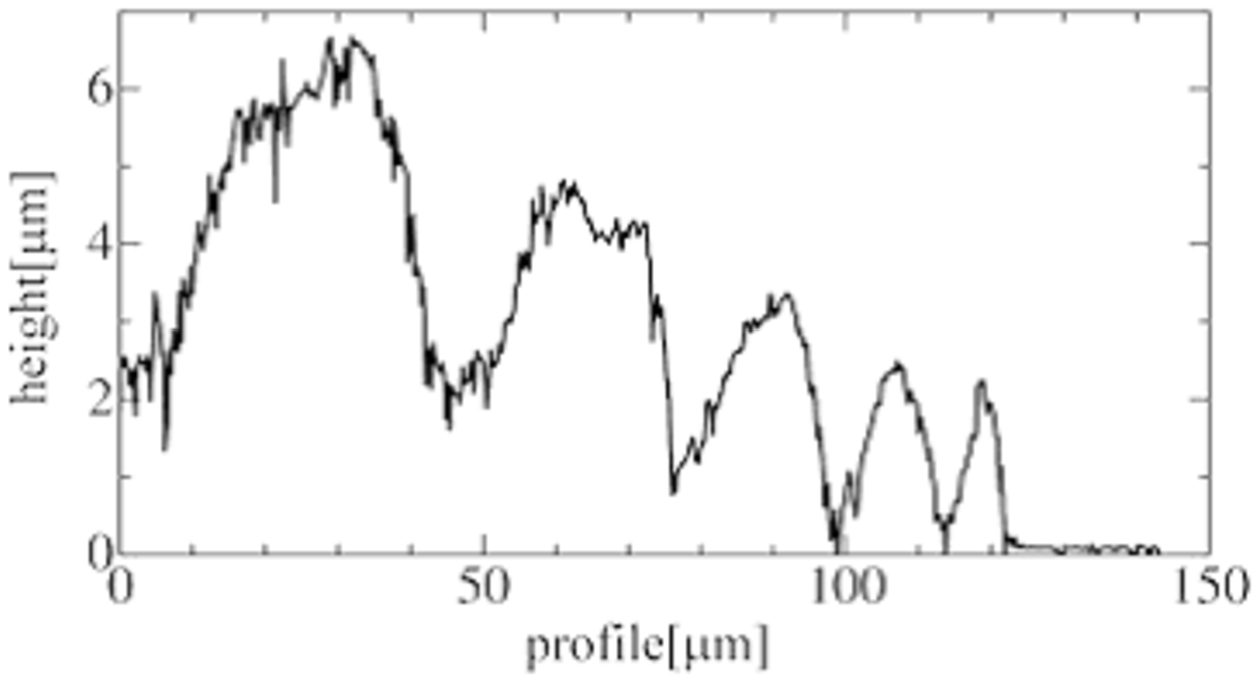}\\
(a) \hspace{3cm} (b)
\caption{Collision of two domains in the periodic growth mode
($H =$ 65 \%, $\rho =$ 1.5 mg/cm$^{2}$, $T =$ 30  $^{\circ}$C). 
}
\label{fig:h-col}
\end{center}
\end{figure}

%
Figure \ref{fig:h-col} shows the boundary of two domains
spreading in the direction of the white arrows.
%
The panel (b) in this figure shows the height profile measured 
along the dotted line in (a).
%
It is found from this figure that the peak height and the pitch
of the domain in the periodic growth decrease as the two domains approach.
%
Considering that the peak height is given
as a monotonic increasing function of 
the thickness of the solution, 
this result indicates that the solution thickness becomes gradually small.
%
The important point here is 
the solution between the two domain fronts.

\begin{figure}[!p]
\begin{center}
\includegraphics*[width=5cm]{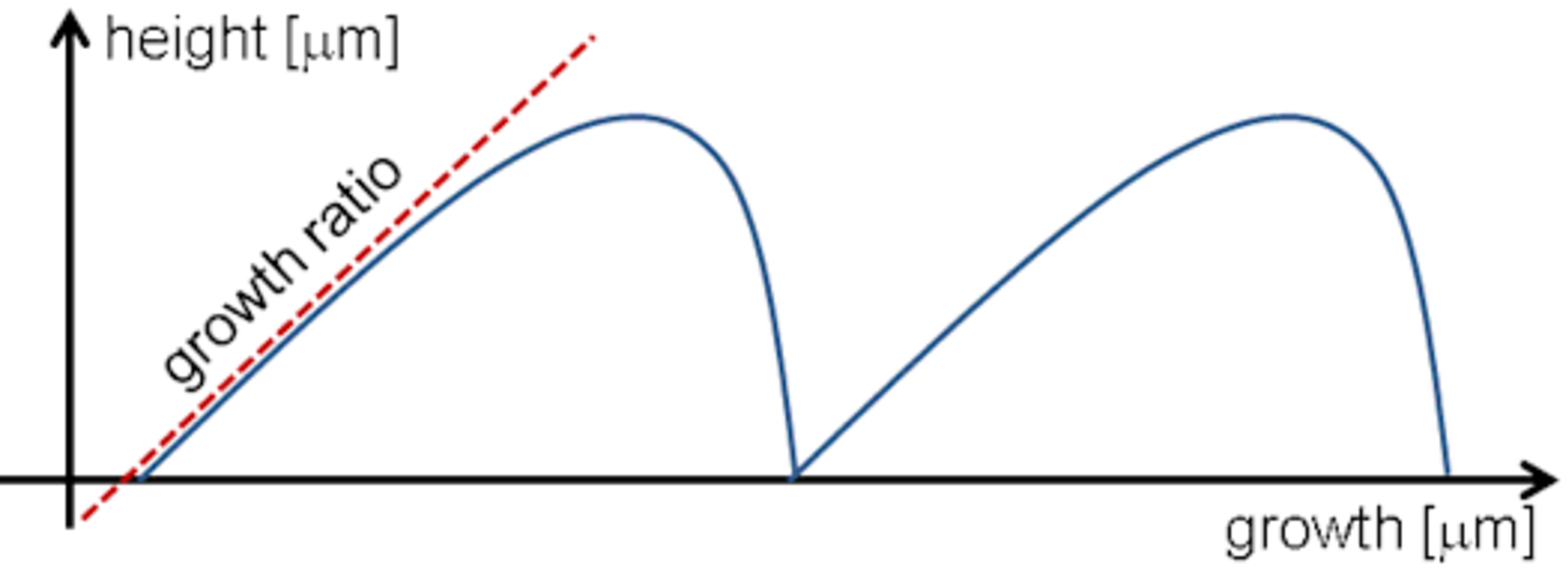}
\includegraphics*[width=5cm]{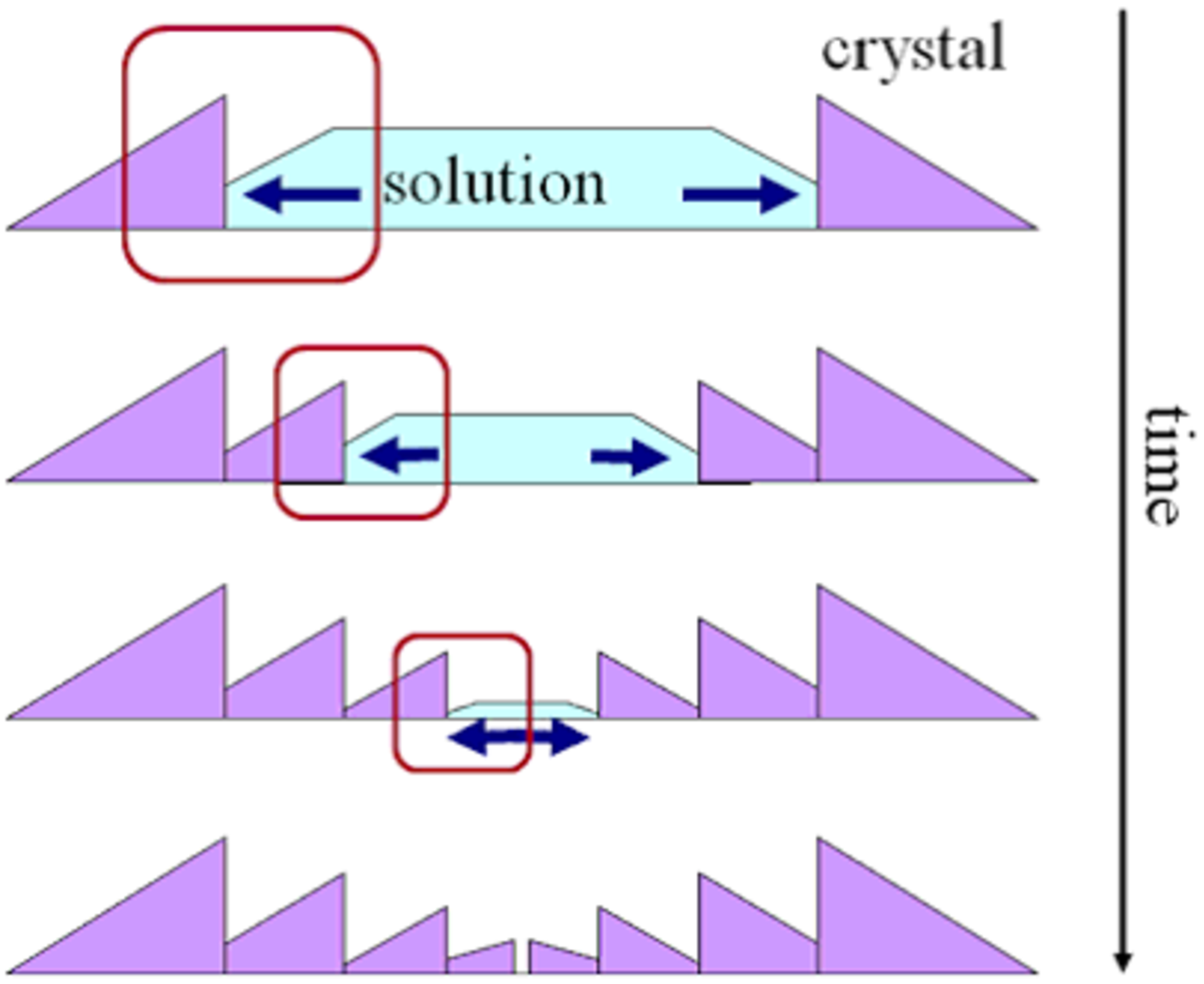}\\
(a) \hspace{3cm} (b)
\caption{(a) Definition of growth ratio for the crystal height.
(b) Self-similar periodic growth.
}
\label{fig:h-sch}
\end{center}
\end{figure}

\begin{figure}[!p]
\begin{center}
\includegraphics*[width=5cm]{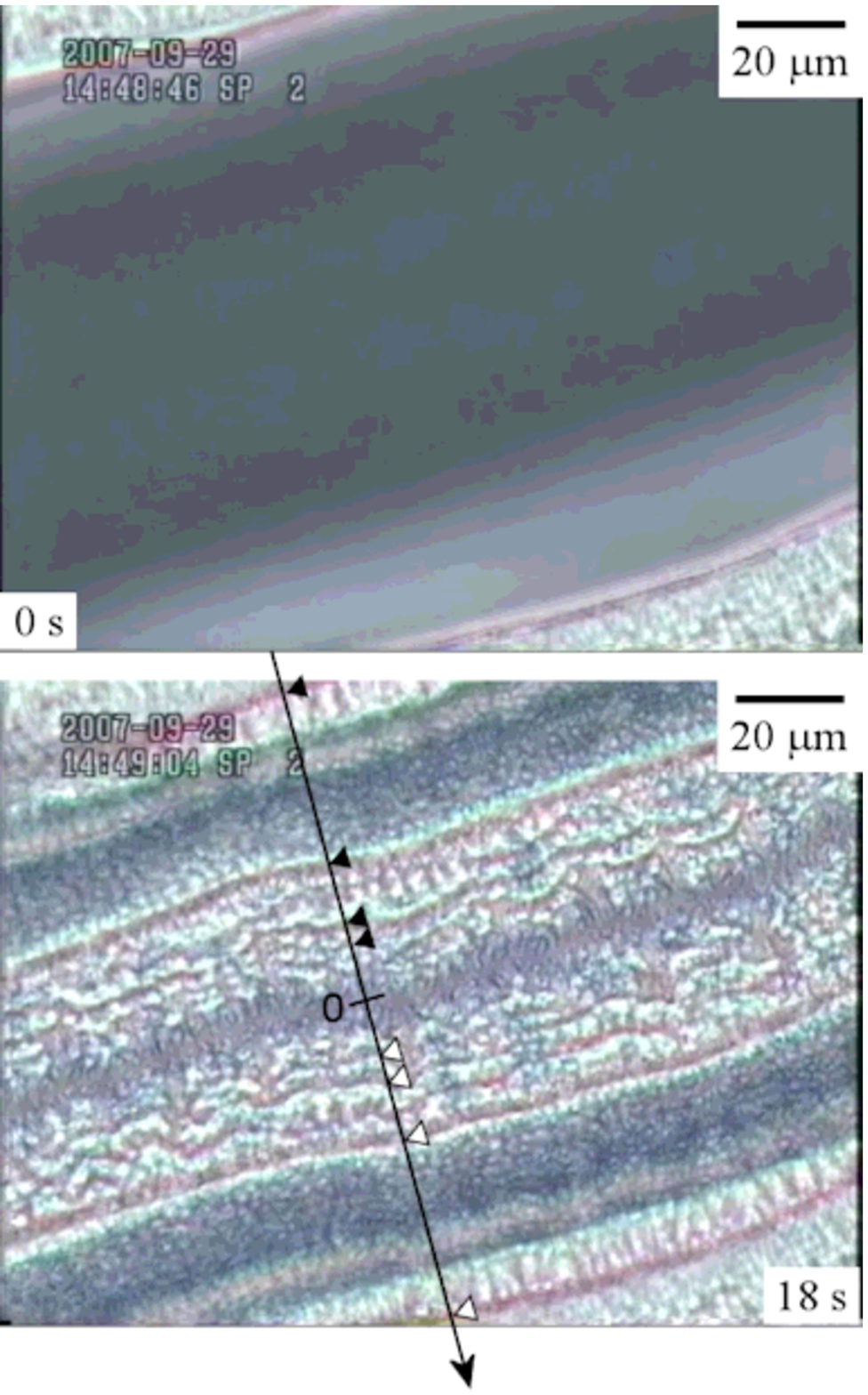}
\includegraphics*[width=5cm]{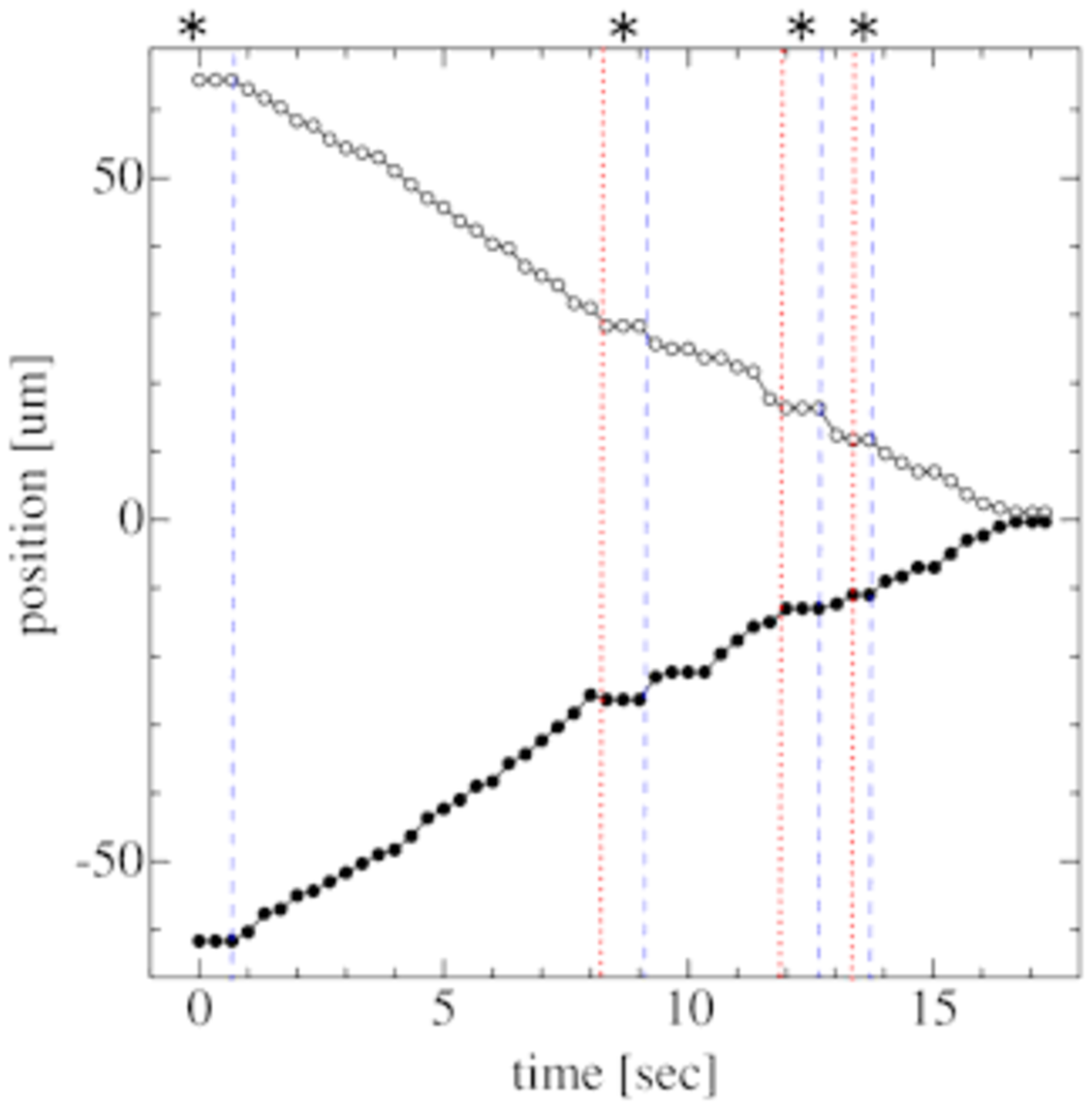}\\
(a) \hspace{3cm} (b)
\caption{Collision of two domains 
($H =$ 50 \%, $\rho =$ 0.5 mg/cm$^{2}$, $T =$ 27  $^{\circ}$C). 
Pitch of periodicity for each domain is represented by 
$\triangle$ and $\blacktriangle$.
The symbol $\ast$ in (b) shows the suspension period.
}
\label{fig:col-sync}
\end{center}
\end{figure}

%
The growth ratio for the height in the periodic growth can be defined 
as the positive slope schematically depicted in Fig. \ref{fig:h-sch}(a).
%
From the panels (b) of Figs. \ref{fig:h-p} and \ref{fig:h-col},
the growth ratio for height is estimated as 0.24 in both cases.
%
Then, we conjecture that 
the growth ratio has less influence on the solution thickness 
for fixed $H$ and $T$.
%
In crystallization, the thickness of the solution surrounded by the domain fronts 
becomes gradually small as the fronts approach each other.
%
This is because the solution flows to the domain fronts 
and is crystallized there.
%
These dynamical features bring about emergence 
of a self-similar periodic growth as shown in Fig. \ref{fig:h-sch}(b).
%
In the self-similar periodic growth, 
a peak formation is repeated with different scales
depending on the solution thickness.
%
In actuality, Fig. \ref{fig:h-col}(a) is one example.
%
As another feature, it is noted that the advance and suspension periods
can be synchronized for the two domain fronts approaching each other.
%
Figure \ref{fig:col-sync} shows the case.
%
This synchronization suggests existence of a non-local interaction 
between the two domain fronts via the thin solution film.

\begin{figure}[!p]
\begin{center}
\includegraphics*[width=5cm]{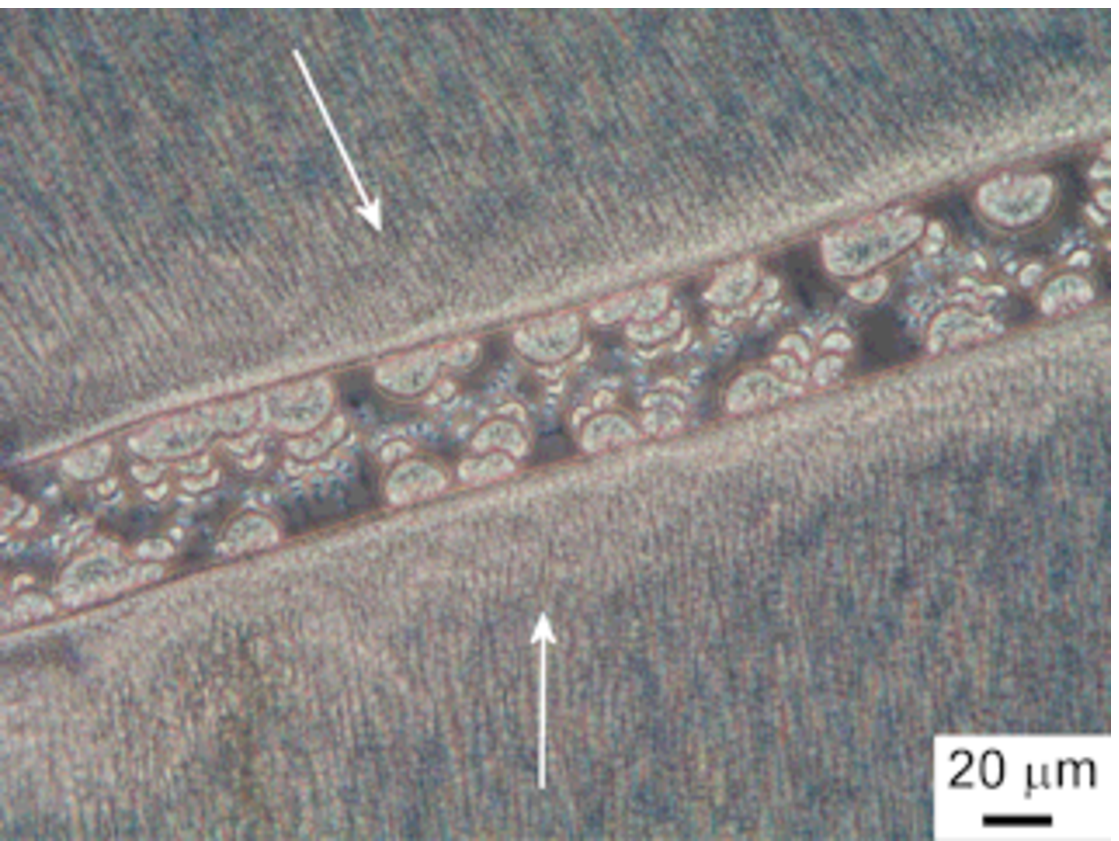}
\includegraphics*[width=5cm]{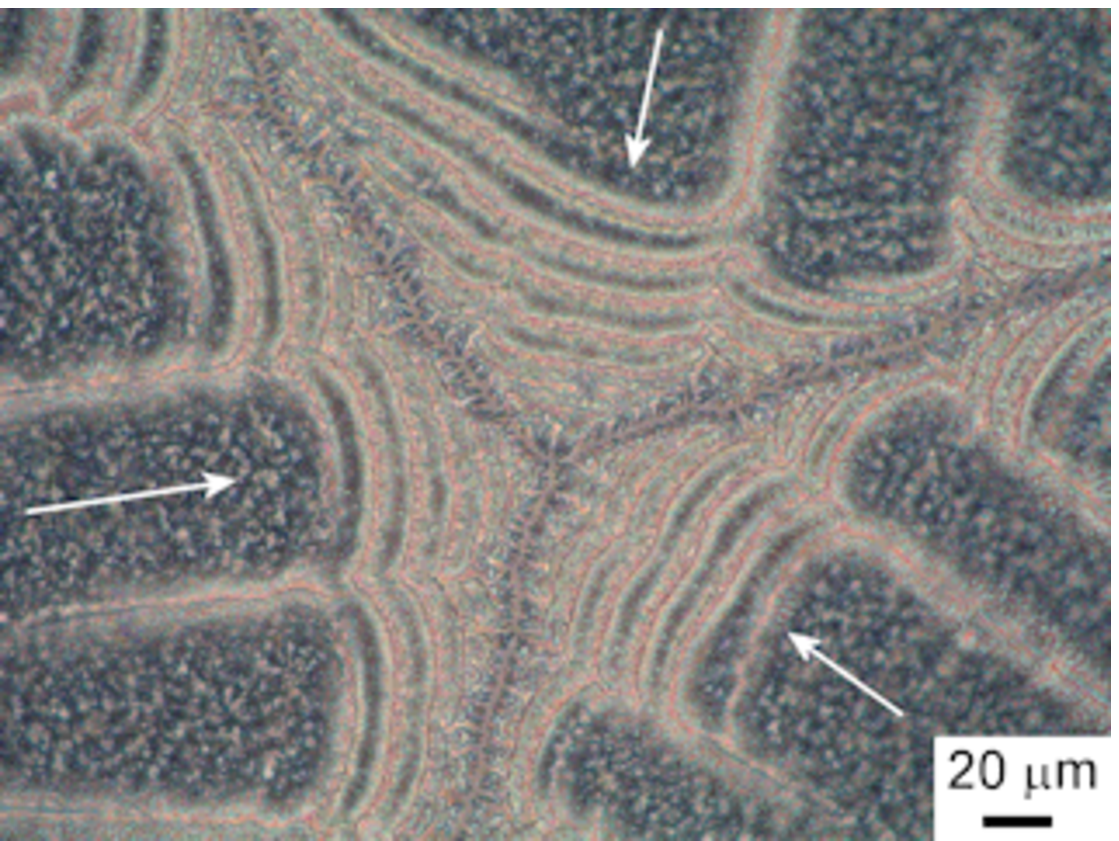}\\
(a) \hspace{3cm} (b)
\caption{Threshold-sensitive transition from the uniform to the periodic growth modes 
($H =$ 50 \%, $\rho =$ 0.5 mg/cm$^{2}$, $T =$ 30  $^{\circ}$C). }
\label{fig:u-p}
\end{center}
\end{figure}

\begin{figure}[!p]
\begin{center}
\includegraphics*[width=5cm]{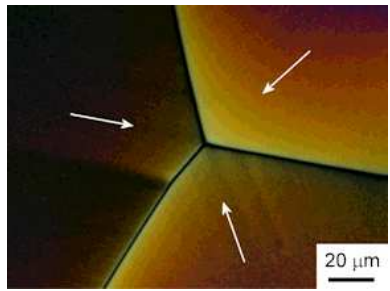}\\
\caption{Collision of three domains 
($H =$ 40 \%, $\rho =$ 1.0 mg/cm$^{2}$, $T =$ 30 $^{\circ}$C). 
In this case, the threshold-sensitive transition does not occur
due to the low humidity.
}
\label{fig:col-low}
\end{center}
\end{figure}

\begin{figure}[!p]
\begin{center}
\includegraphics*[width=5cm]{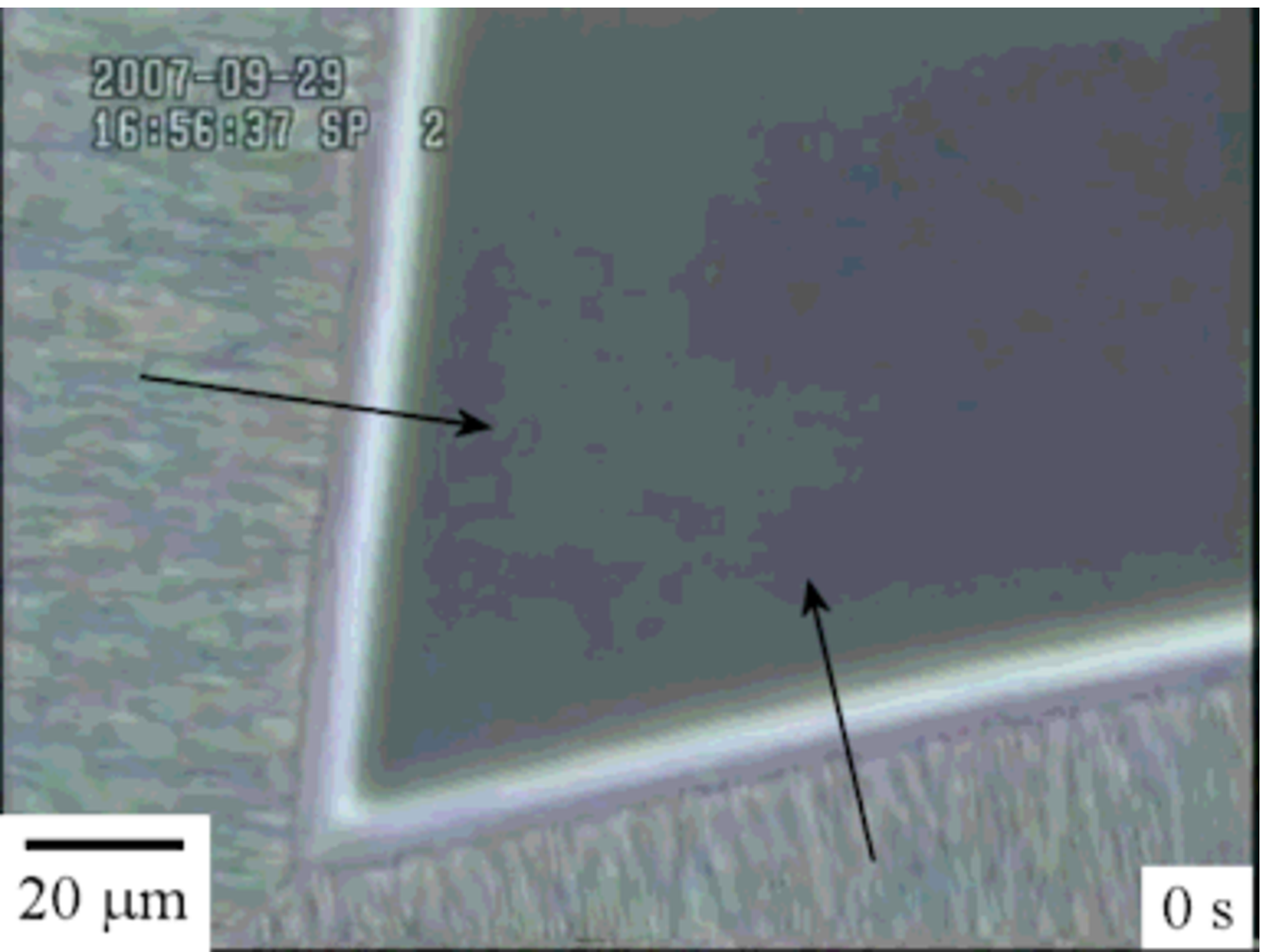}
\includegraphics*[width=5cm]{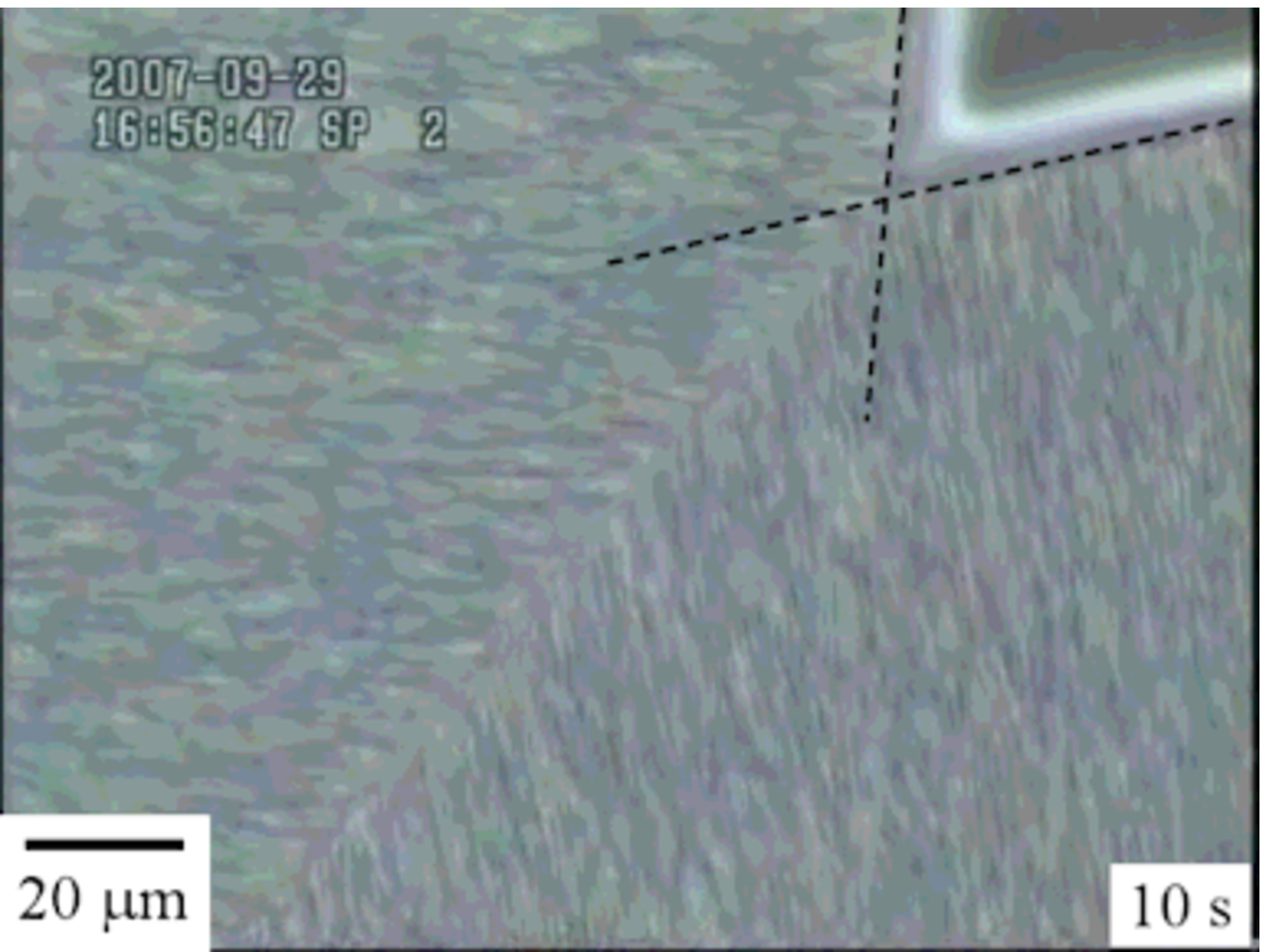}\\
(a) \hspace{3cm} (b)
\caption{Collision of two domains 
($H =$ 50 \%, $\rho =$ 0.5 mg/cm$^{2}$, $T =$ 30  $^{\circ}$C). 
In this case, the threshold-sensitive transition does not occur
since the solution is not surrounded by the domain fronts.
}
\label{fig:col-theta}
\end{center}
\end{figure}

%
Furthermore, it is found that transition from the uniform to the periodic modes
occurs in collision as shown in Fig. \ref{fig:u-p}.
%
Such a transition was frequently observed 
when (i) humidity for the uniform growth is relatively high ($\gtrsim$ 50\%)
and (ii) the solution region is closed for being surrounded by domain fronts.
%
Hence, this transition does not occur if the above two conditions
are not satisfied.
%
Actually, Fig. \ref{fig:col-low} shows the case 
where the condition (i) is not satisfied: 
a lower humidity condition ($\approx$ 40 \%), 
even if the condition (ii) is satisfied.
%
And as shown in Fig. \ref{fig:col-theta}, this transition 
does not occur when the solution region is open 
(to the upper right side in this figure), even if the condition (i) is satisfied.
%
Therefore, it is considered that this dynamical transition occurs 
when the solution thickness reaches the threshold 
as the thickness becomes small by the solution flow in crystallization.

%

%

%

%

%

%

\section{A toy model}
\label{sec:scenario}

%
The essential factor to be considered 
for this pattern forming system 
is flow of the thin solution film with its thickness threshold.
%
However, detailed hydrodynamical properties of the thin solution film 
in crystallization seems to be quite difficult to identify.
%
In this paper, we construct a toy model 
based on the above experimental results, 
so that we qualitatively grasp the dynamical properties 
of the uniform and the periodic growth modes.

\begin{figure}[!p]
\begin{center}
\includegraphics*[width=7.0cm]{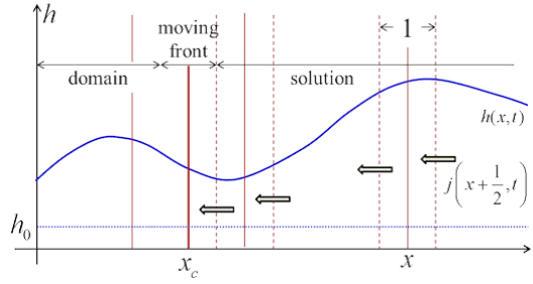}
\caption{The setup for the toy model.}
\label{fig:model}
\end{center}
\end{figure}

%
As shown in Fig.\ref{fig:model}, the two dimensional solution film 
spreading along the $x$ direction is considered.
%
In this model, time and space are discretized 
by a time step $\Delta t$ 
and by using a staggered grid, where the unit grid size is $\Delta x$, 
respectively.
%
Here, we set $\Delta t$ = 1 and $\Delta x$ = 1.
%
$x_{c}$ shows the growth front 
and is assumed to move in the direction of positive $x$.
%
As the initial condition, 
crystallization starts at $x_{c}$ = 0 when $t = 0$.
%
The regions $0 \leq x < x_{c}(t)$ and $x_{c}(t) < x$ 
correspond to the crystal domain and the solution film, respectively.

%
The solution thickness and the domain height are represented by $h(x, t)$ in common.
%
If we consider only the solution flow $j$ 
for change of $h$ in time, 
$h$ satisfies the equation of continuity,
%
\begin{equation}
  h(x, t + 1) - h(x, t) = 
  \begin{cases}
    - j \left(x + \frac{1}{2}, t \right)
      +j \left(x - \frac{1}{2}, t \right) & (x \geq x_{c}(t)), \\
    0 & (x < x_{c}(t)).
  \end{cases}
  \label{eqn:h}
\end{equation}
%
The direction of the flow in the present case is always negative ($j < 0$).
%
When $h$ becomes lower than the threshold $h_{0}$ ($x \geq x_{c}(t)$), 
the flow stops ($j = 0$).
%
At the domain front ($x = x_{c}$), 
the flow $f_{c}$ is caused by the crystallization.
%
It is considered that $f_{c}$ depends on 
the domain height in contact with the growth front: $h(x_{c} - 1, t)$.
%
And humidity is considered to increase $f_{c}$,
since the evaporation rate becomes low and 
the fluidity of the solution enhances.
%
Now, $f_{c}$ is assumed to be given as
%
\begin{equation}
  f_{c} = a h(x_{c} - 1, t) \theta \left( h(x_{c}, t) - h_{0} \right).
  \label{eqn:f}
\end{equation}
%
Here $a$ is a humidity-dependent positive parameter, and 
$\theta$ is the step function: $\theta(x)$ = 1 $(x \geq 0)$, 0 $(x < 0)$.
%
The solution flow $j$ is induced by $f_{c}$, where $x > x_{c}$.
%
It is considered that $j(x+1/2)$ has a monotonically increasing dependence on $h(x)$ 
and becomes gradually decreasing 
as the distance between $x$ and $x_{c}$ becomes large.
%
Now, we simply assume the flow $j$ as
%
\begin{equation}
  j \left(x + \frac{1}{2}, t \right) =
  \begin{cases}
    - f_{c} & (x = x_{c}), \\
    - h(x, t) \exp \left( - \gamma (x - x_{c})\right) f_{c} & (x > x_{c}), \\
  \end{cases}
  \label{eqn:j}
\end{equation}
%
where $\gamma$ is a positive constant.
%
For the motion of growth front $x_{c}(t)$, 
its speed in the advance period, $v_{a}$, 
is assumed to be almost constant based on our experimental results.
%
And the front comes into the suspension period when $h(x_{c}) \leq h_{0}$.
%
During the suspension period, the film thickness becomes so small 
that the growth speed is quite slow as shown in Fig. \ref{fig:vel-u-b}.
%
And the growth front gradually moves to the region 
where the film thickness is bigger than the threshold.
%
Here, the time for the suspension period is represented by $\tau_{s}$.
%
Taking the discreteness of this model into account,
the motion of the growth front can be described 
by using the floor function as
%
\begin{equation}
  x_{c}(t) = 
  \left\lfloor \sum_{t^{\prime}=0}^{t} \left\{
    v_{a} \theta^{\prime}
    + \tau_{s}^{-1} \left( 1 - \theta^{\prime} \right) \right\} \right\rfloor,
  \label{eqn:x}
\end{equation}
where $\theta^{\prime} = \theta \left( h(x_{c}(t^{\prime}), t^{\prime}) - h_{0} \right)$, 
and $\left\lfloor x \right\rfloor$ is the floor function 
which gives the maximum integer smaller than $x$.

%

\begin{figure}[!p]
\begin{center}
\includegraphics*[width=7.0cm]{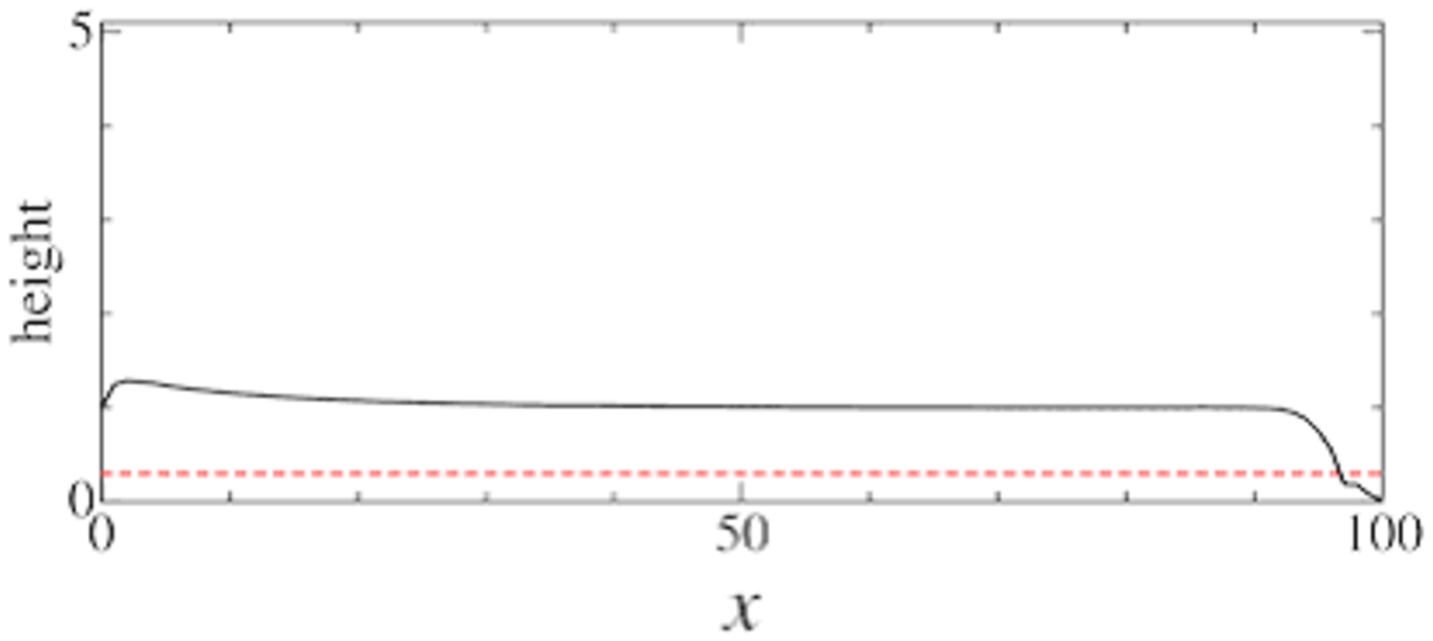}
\hspace{1cm}
\includegraphics*[width=7.0cm]{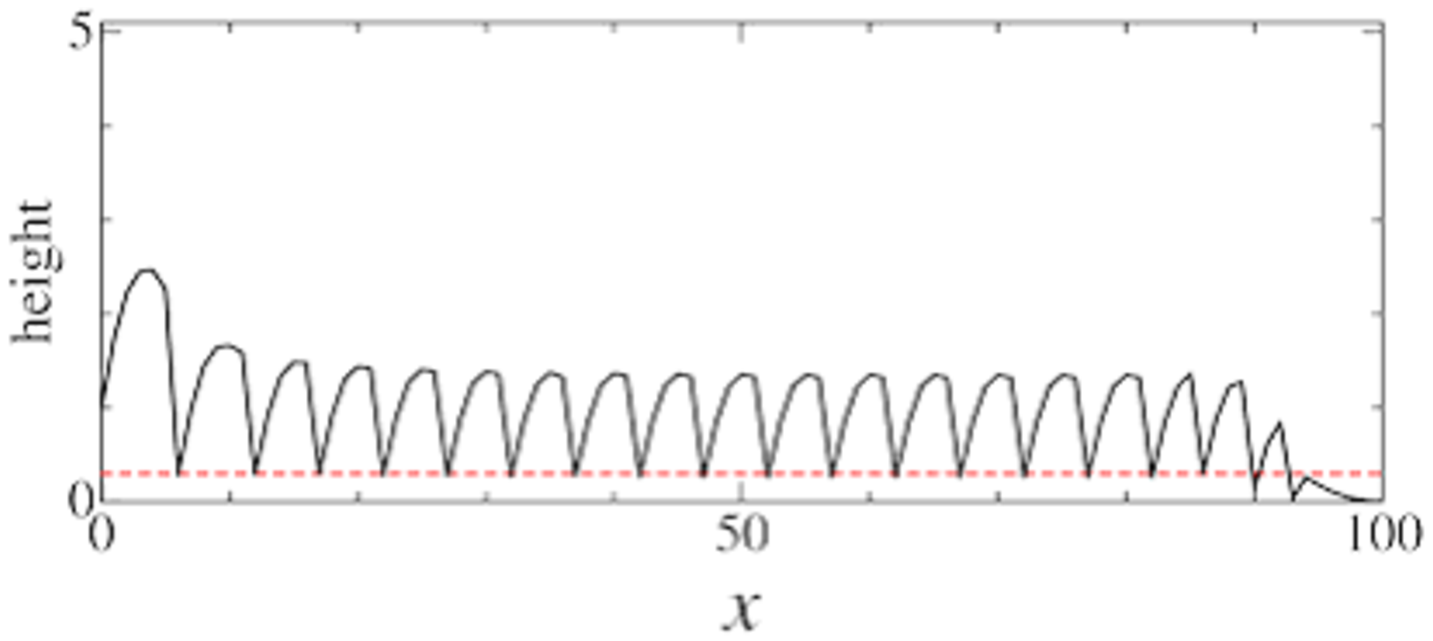}\\
(a)\hspace{7cm}(b)
\caption{The height profiles after time evolution by the model
until $x_{c}=x_{b}$. $\gamma$ = 0.05. $a$ = (a) 0.005 and (b) 0.015.
The red broken line shows the threshold $h_{0}$.}
\label{fig:transition}
\end{center}
\end{figure}
\begin{figure}[!p]
\begin{center}
\includegraphics*[width=7.0cm]{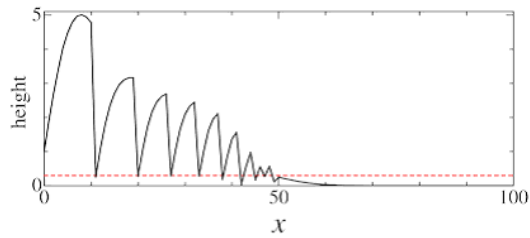}
\caption{The self-similar peak formation obtained by the model.
$\gamma$ = 0.005 and $a$ = 0.018.
The red broken line shows the threshold $h_{0}$.}
\label{fig:self-similar}
\end{center}
\end{figure}
\begin{figure}[!p]
\begin{center}
\includegraphics*[width=7.0cm]{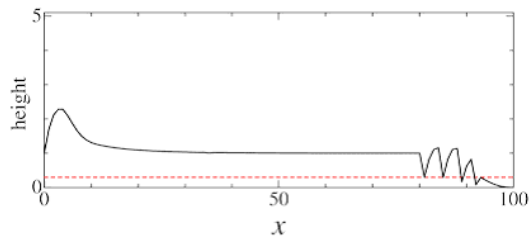}
\caption{The threshold-sensitive dynamical transition from the uniform 
to the periodic growth modes. $\gamma$ = 0.05 and $a$ = 0.014.
The red broken line shows the threshold $h_{0}$.}
\label{fig:thre-sen}
\end{center}
\end{figure}

%
For numerically solving eqs.(\ref{eqn:h})-(\ref{eqn:x}), 
we set the system size as $x_{c}(0) = 0 \leq x \leq x_{b}$.
%
The results shown here are the cases where 
$v_{a}$ = 0.02, $\tau_{s}^{-1}$ = 0.01, $h_{0}$ = 0.3, and $x_{b}$ = 100.
%
As the initial condition, $h(x,0)$ = 1 for $x < x_{b}$.
%
And we set $h(x_{b}, t)$ = 0 for the boundary condition.
%
Figure \ref{fig:transition} shows the height profiles after $x_{c}$ reaches $x_{b}$ 
for different values of $a$.
%
The panels (a) and (b) correspond to the uniform and the periodic growth modes, 
respectively.
%
It is found from Fig. \ref{fig:self-similar} that this model is able to reproduce 
the self-similar peak formation (see Figs. \ref{fig:h-col} and \ref{fig:h-sch} for reference).
%
Furthermore, threshold-sensitive transition from the uniform to the periodic growth modes 
is reproducible as shown in Fig. \ref{fig:thre-sen}
(see Fig. \ref{fig:u-p} for reference).

%

%

%

\section{Discussion}
\label{sec:discuss}

\subsection{Metastability and humidity dependence of the solution}
\label{sec:metastable}
%
When the supersaturated solution 
without a crystal domain was set in $H$ $\approx$ 0 (less than 5\%), 
crystal growth did not occur over two weeks.
%
Here the zero humidity environment was realized 
by putting a large amount of silica gel together in a closed box.
%
This metastable property of the solution 
is consistent with the report by Gunn\cite{Gunn09}.
%
Furthermore, we found that the crystal domain started to grow
when the solution was put in a humidity 
between 10 \% and 70 \% even after the two weeks.
%
And the domain growth stopped when we set $H$ $\approx$ 0 again.

%
From these observations, the following features
about humidity dependence of the supersaturated solution are suggested.
%
(i) In the ``glassy'' solution under the zero humidity condition, 
ascorbic acid is immovable, 
and then the crystal growth stops or does not occur.
%
(ii) Rise in $H$ brings about 
increase of water content in the solution film, 
and also decline of water evaporation rate in crystallization.
%
(iii) Increase of water content in the solution film 
enhances the mobility of ascorbic acid, 
and the crystallization becomes possible.
%
(iv) Decline of water evaporation rate causes fluidization of the solution
near the growth front.
%
The fluidization also enhances the mobility of ascorbic acid.
%
According to the report by Ito et al.\cite{Ito03}, 
monocrystal was formed in a high humidity case ($\sim$ 80 \%).
%
This is because the solution released in crystallization remains
around the domain front 
and fluidizes the supersaturated solution.
%
Then, fluidity of the solution around the domain becomes high, 
and crystallization occurs only at a nucleation point
instead of spreading of the crystal domain.

\subsection{Flow of residual solution in a high humidity condition}

\begin{figure}[!p]
\begin{center}
\includegraphics*[width=10cm]{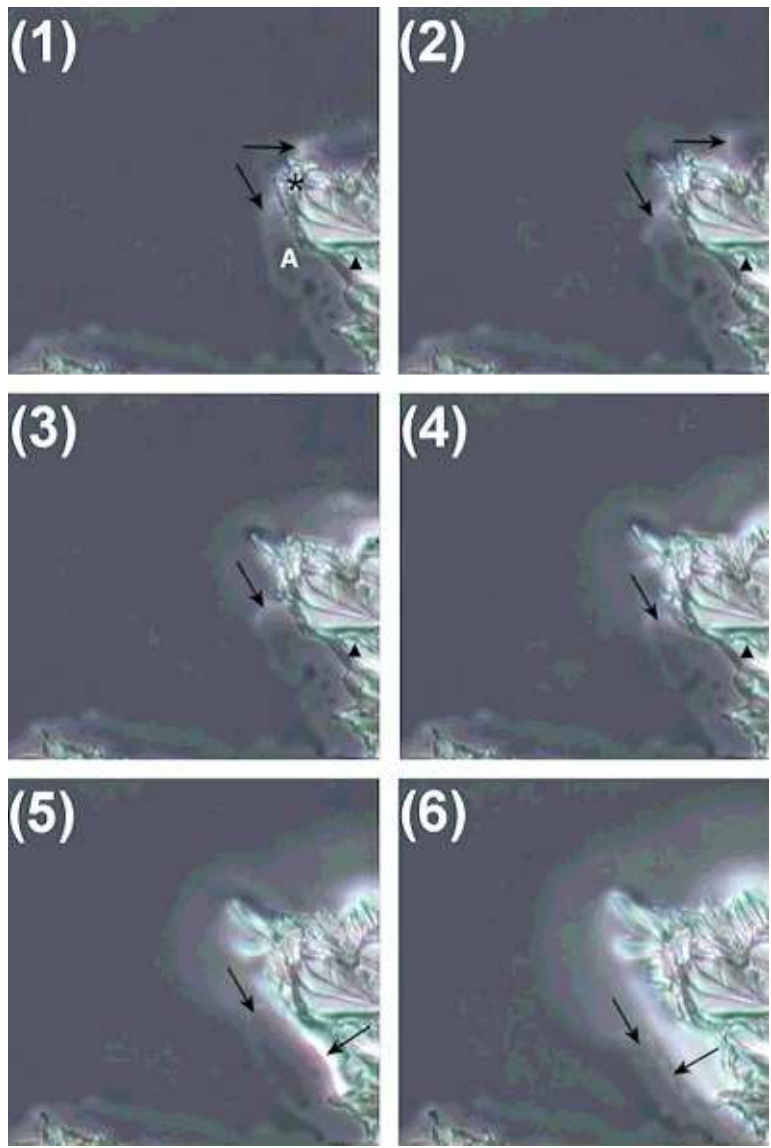}\\
\caption{Flow of the residual solution at the beginning 
of the advance period. 100 $\mu$m $\times$ 100 $\mu$m
($H =$ 75 \%, $\rho =$ 0.5 mg/cm$^{2}$, $T =$ 27 $^{\circ}$C).
Time interval is 2/3 secs.
}
\label{fig:fl-b-out2}
\end{center}
\end{figure}

%
Figure \ref{fig:fl-b-out2} shows a time sequence of the crystal growth 
and the flow of the residual solution released from the crystal domain.
%
This sequence shows the beginning of the advance period 
in the periodic growth mode.
%
In the panel (1), the crystal grows at the point ``$\ast$''.
%
Former crystallization brings about emergence of the region A 
in which the solution does not exist.
%
The black arrows in each panel point at the boundary 
of the regions whether the solution exists or not.
%
As time proceeds, the boundaries move so that 
the region A shrinks, since the released solution in crystallization 
spreads over the region A.
%
As an evidence for the fact that there is no solution in the region A, 
crystal growth does not occur from the side pointed by $\blacktriangle$ 
shown in the panels (1)-(4).
%
In the panel (5), however, the crystal grows from the inside of the domain.
%
And the solution released by crystallization flows out in the region A
as shown in the panel (6).

%
%

\begin{figure}[!p]
\begin{center}
\includegraphics*[width=10cm]{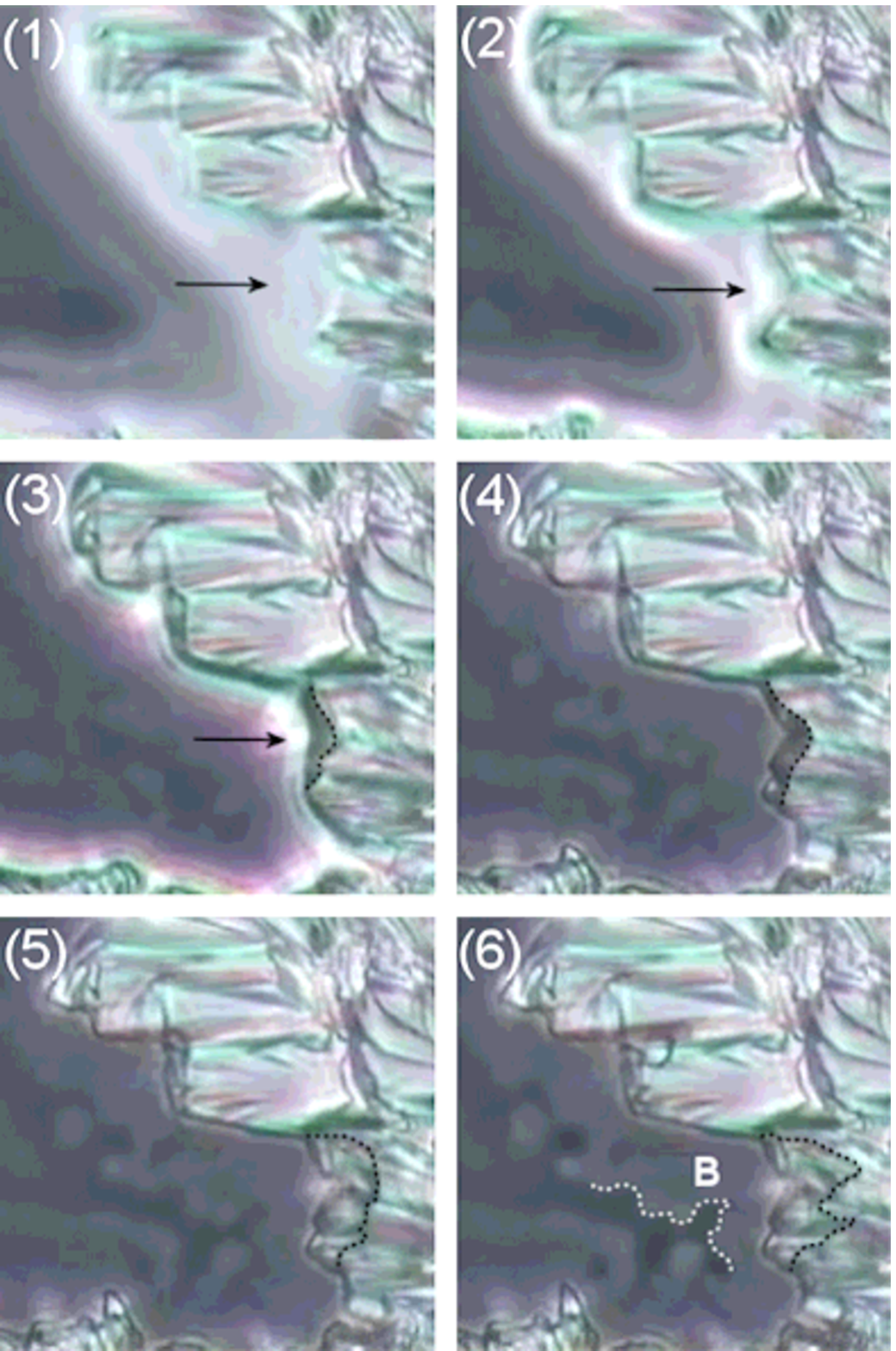}\\
\caption{Flow of the residual solution just before
the suspension period. 100 $\mu$m $\times$ 100 $\mu$m
($H =$ 75 \%, $\rho =$ 0.5 mg/cm$^{2}$, $T =$ 27 $^{\circ}$C). 
Time interval is 2/3 secs.
}
\label{fig:fl-b-in}
\end{center}
\end{figure}

%
On the other hand, Fig. \ref{fig:fl-b-in} shows a time sequence
of the solution flow directed to the domain front, 
so that the solution was absorbed in the domain.
%
These snapshots show the case just before the suspension period.
%
The black arrows in the panels (1)-(3) 
indicate the direction of the solution flow.
%
And the black dotted curves in the panels (3)-(6) show 
the interface between the solution and the air.
%
As time proceeds, the interface moves inside the domain.
%
Therefore, it is found that 
there exist some spacings between the crystals in the domain,
and that the solution remains in the spacings.
%
Due to existence of the residual solution, 
the crystal growth can occur irregularly from
the inside of the domain 
as shown in the panels (5) and (6) of Fig. \ref{fig:fl-b-out2}.
%
Note that the irregularity shown in Fig. \ref{fig:vel-b} is the same.
%
In the region B between the domain front and the white dotted curve,
the solution does not exist.
%
This is because the supersaturated solution in the region B is dissolved
by the residual solution and is totally absorbed in the spacings.
%
This absorption seems to be caused by capillary force as pointed out by Uesaka.
%
These features can be remarkably observed in not only the branching mode, 
but also the periodic mode in a relatively high humidity ($\approx$ 70 \%).

%

%

\subsection{Relevance to the previous studies}

%
Our present scenario is essentially different from that by Uesaka et al.
%
In their conclusion, the dynamical transition 
between the uniform and the periodic modes depends on 
whether the local fluidization of the supersaturated solution film 
occurs or not.
%
However, our observation reveals that the fluidized region 
is formed in both modes.
%
Therefore, formation of the fluidized region is not the essential factor 
for the dynamical transition.
%
The thickness threshold of the solution film is essential.
%
Moreover in their scenario, the local dynamics in the vicinity of the growth front
was focused on.
%
Actually they constructed a phase field model with reaction-diffusion type dynamics 
at the growth front \cite{Uesaka02}.
%
In their model, state change of ascorbic acid and 
formation of spacings between crystals were considered.
%
However, it is suggested from our observation 
that there exists the non-local interaction coupled with domain growth motion.
%
Then in our model, the dynamics of the solution flow is mainly focused on.

%
%

%
Regarding hydrodynamics of thin liquid film, 
many experimental and theoretical studies have been carried out 
\cite{Sharma93, Williams82, Craster09, Oron97, deGennes04}.
%
This pattern forming phenomenon can be considered as 
viscous fingering with variable thickness and the threshold.
%
And the dynamical properties in this system seems to be related to 
that in drying water film \cite{Merzel98, Lipson98} 
and to the ring-banded spherulites grown from the thin film polymeric solution
\cite{Wang07, Wang08}.

\section{Summary and conclusion}
\label{sec:conclusion}

%
The dynamical properties and the humidity dependences 
of the uniform and the periodic growth modes
of ascorbic acid crystal domain from thin solution film are summarized as follows.

\textit{Fluidity of the solution film}:
%
Flow of the supersaturated solution is caused by crystallization 
in the direction to the domain front.
%
Especially in the periodic growth, 
the flow stops in accordance with the stop of the crystal growth.
%
The solution flow decreases the film thickness.
%
The two characteristic solution regions can be formed 
in the vicinity of the domain front: 
(1) the local high fluidity region and 
(2) the region having smaller thickness.
%
These regions can be observed in both the uniform and the periodic growth modes.

\textit{Humidity dependence}:
%
In the zero humidity condition, the crystal growth does not occur or stops.
%
%
In the case of 50 \% $\lesssim H \lesssim$ 75 \%, 
the front speed becomes almost constant regardless of the growth modes.
%
When $H \gtrsim$ 70 \%, 
it is remarkable that the solution released in crystallization remains
in front of the domain front.
%
The residual solution exhibits the bidirectional flows 
and brings about irregularity in the periodic growth.

\textit{Thickness of the solution film}:
%
As the thickness of the solution film becomes small, 
the front speed becomes low.
%
This brings about a thickness threshold for the domain growth.
%
The threshold causes a dynamical transition from the uniform to the periodic modes.
%
The growth ratio for the domain height is independent of the film thickness.
%
Then, by considering the threshold together, 
the pitch of the rings in the periodic growth depends on the film thickness, 
and a self-similar periodic growth is realized.
%
In the periodic growth during collision of domain fronts, 
the fronts motions can be synchronized.

%
%

%
In conclusion, we have reported the dynamical properties and humidity dependence 
of the domain front motion and the solution flow in crystallization 
of ascorbic acid crystal from its thin solution film.
%
The crystal growth involves the solution flow.
%
Humidity enhances the fluidity of the solution 
due to the lowering of the evaporation rate.
%
There exists the threshold for thickness of the solution film.
%
Below the threshold, the crystal growth almost stops.
%
As a result, the threshold-sensitive transition 
from the uniform to the periodic growth modes is realized.

\section*{Acknowledgement}

%
We thank Prof. S. Tanaka (Hiroshima University) 
for supplying us the latex beads with fruitful discussions.
%
We also acknowledge 
Mr. S. Ishii, Dr. K. Seki, Prof. Y. Tabe, and Prof. S. Ishiwata (Waseda University)
for some measurements regarding fluidity of the solution film 
and for helpful advices.
%
We are also grateful to Prof. K. Taguchi(Hiroshima University)
for the related information of polymer crystallization.
%
This work was supported by MEXT KAKENHI(Priority Areas,
``Creation of Nonequilibrium Soft Matter Physics''),
a Waseda University Grant for Special Research Projects (2012B-150), 
and JSPS KAKENHI(Challenging Exploratory Research) 
No. 25610110.


\newpage
%
%
%

%

\clearpage

%
%

%
%

%

\newpage

%
%

%

\clearpage

%
%

%

\clearpage

%
%

%
%

%

\clearpage

%
%

%

\clearpage

%
%

%
%

%

\newpage

%
%

%
%

%

\clearpage

%
%

%
%
%
%
%
\newpage

%

\clearpage

%


\begin{thebibliography}{99}
\bibitem{Epstein96}%
  I. R. Epstein, K. Showalter: J. Phys. Chem. 100 (1996) 13132-13147.
%
\bibitem{Epstein98}%
  I. R. Epstein, J.A. Pojman: \textit{An introduction to nonlinear chemical dynamics: oscillations, waves, patterns, and chaos}
  (Oxford University Press, USA, New York, 1998).
%
\bibitem{Itoh99}%
  H. Itoh, J. Wakita, T. Matsuyama, M. Matsushita, J. Phys. Soc. Jpn., 68 (1999) 1436-1443
%
\bibitem{Wakita01}%
  J. Wakita, H. Shimada, H. Itoh, T. Matsuyama, M. Matsushita, J. Phys. Soc. Jpn., 70 (2001) 911-919
%
\bibitem{Henisch05}%
  H.K. Henisch, \textit{Crystals in gels and Liesegang rings} (Cambridge University Press, New York, 2005).
%
\bibitem{Terada48}%
  T. Terada, \textit{Shizenkai no shimamoyou} (Iwanami shoten, Tokyo, 1948) [in Japanese].
  http://www.aozora.gr.jp/cards/000042/files/2354\_13803.html.
%
\bibitem{Wang07}%
  Z. Wang, Z. Hu, Y. Chen, Y. Gong, H. Huang, T. He, Macromolecules, 40 (2007) 4381-4385.
%
\bibitem{Wang08}%
  Z. Wang, G. Alfonso, Z. Hu, J. Zhang, T. He, Macromolecules, 41 (2008) 7584-7595.
%
\bibitem{Gunn09}%
  Gunn, E. (2009). Small molecule banded spherulites. Ph.D. 3377298, University of Washington
%
\bibitem{Shtukenberg11}%
  A. Shtukenberg, E. Gunn, M. Gazzano, J. Freudenthal, E. Camp, R. Sours, E. Rosseeva, B. Kahr, ChemPhysChem, 12 (2011) 1558-1571.
%
\bibitem{Shtukenberg12}%
  A.G. Shtukenberg, Y.O. Punin, E. Gunn, B. Kahr, Chem. Rev., 112 (2012) 1805-1838.
%
\bibitem{Iwamoto84}%
  K. Iwamoto, S.I. Mitomo, M. Seno, J. Colloid Interface Sci., 102 (1984) 477-482
%
\bibitem{Fukunaga96}%
  K. Fukunaga:
  Chemical Education \textbf{44} (1996) 548-549  [in Japanese].
%
\bibitem{Uesaka02}%
  H. Uesaka, R. Kobayashi, J. Cryst. Growth, 237-239 (2002) 6.
%
\bibitem{Uesaka03}%
  H. Uesaka, R. Kobayashi, T. Yamaguchi, \textit{Analysis of Nonlinear Phenomena}
  (RIMS Kokyuroku vol.1313) (2003) 25-35  [in Japanese].
%
\bibitem{Ito03}%
  M. Ito, M. Izui, Y. Yamazaki, M. Matsushita, J. Phys. Soc. Jpn., 72 (2003) 1384-1389
%
\bibitem{Paranjpe02}%
  A. Paranjpe, Phys. Rev. Lett., 89 (2002) 75504.
%
\bibitem{Yamazaki09}%
  Y. Yamazaki, H. Yoshino, M. Izui, Y. Sato, M. Matsushita, J. Phys. Soc. Jpn., 78 (2009) 074001
%
%
\bibitem{Sharma93}%
  A. Sharma, A. T. Jameel, J. Colloid Interface Sci., 161 (1993) 190-208.
%
\bibitem{Williams82}%
  M. B. Williams, S. H. Davis, J. Colloid Interface Sci., 90 (1982) 220-228.
%
\bibitem{Craster09}%
  R. Craster, O. Matar, Reviews of Modern Physics, 81 (2009) 1131.
%
\bibitem{Oron97}%
  A. Oron, S.H. Davis, S.G. Bankoff, Reviews of Modern Physics, 69 (1997) 931-980.
%
\bibitem{deGennes04}%
  P.-G. de Gennes, F. Brochard-Wyart, D. Qu\'{e}r\'{e},
  \textit{Capillarity and Wetting Phenomena: Drops, Bubbles, Pearls, Waves}
  (Springer, New York, 2004).
%
\bibitem{Merzel98}%
  N. Samid-Merzel, S. Lipson, D. S. Tannhauser, Phys. Rev. E, 57 (1998) 2906.
%
\bibitem{Lipson98}%
  S. G. Lipson, N. Samid-Merzel, D. S. Tannhauser, Europhysics news, 29 (1998) 116-120.
%
%
\end{thebibliography}
\end{document}